\documentclass[a4paper,11pt]{article}
\pdfoutput=1 
\usepackage{jheppub} 
                     
\usepackage{booktabs,makecell, bm}
\usepackage[font=small]{caption}
\usepackage{framed}
\usepackage{hyperref}
\usepackage[export]{adjustbox}
\usepackage{amsmath}
\usepackage{lineno}
\usepackage{ytableau}
\usepackage{url}
\usepackage[numbers]{natbib}
\usepackage{amsthm}

\usepackage{xcolor}

\newtheorem{proposition}{Proposition}[section]

\usepackage{physics}
\usepackage{ytableau}
\usepackage{subcaption}

\def\CN{\mathcal{N}}

\title{Probing Non-Graviton Spectra in $\mathcal{N}=4$ SYM via BMN truncation and S-Duality}

\author[a]{Abhijit Gadde,}
\author[a]{Eunwoo Lee,}
\author[a]{Rajat Raj,}
\author[a]{Shivansh Tomar}

\affiliation[a]{Department of Theoretical Physics, \\ Tata Institute
	of Fundamental Research, Homi Bhabha Rd, Mumbai 400005, India}
\emailAdd{abhijit@theory.tifr.res.in}
\emailAdd{eunwoo.lee@tifr.res.in}
\emailAdd{rajat.raj@tifr.res.in}
\emailAdd{shivansh.tomar@tifr.res.in}

\abstract{
The one-loop cohomology of $\CN=4$ super Yang-Mills is conjectured to be isomorphic to the  exact cohomology. As a result, its truncations  are expected to be sub-rings of the exact cohomology. We study the superconformal index restricted over one such truncation known as the BMN truncation. We present a systematic algorithm to compute the BMN index using the method of residues. We compute the BMN  index for SU($N$) $\CN =4$ SYM for \(N= 2, \dots, 6\) in closed form. It is expressed as a rational function of the fugacity. 
A term of the type $(1-x)$ in the denominator indicates the presence of a bosonic generator counted with fugacity $x$. We find a rich and universal set of such terms in the denominator showing an interesting bosonic Fock space in the spectrum of  protected operators. 
This Fock space can not be explained as coming from the non-interacting supersymmetric graviton gas far away from the black hole as in the grey-galaxy solutions because the charges of the bosonic generators are not compatible with those of the supersymmetric gravitons. This suggests a novel micro-structure within the supersymmetric  black hole itself.

We also examine the indices of S‑dual pairs SO($2N+1$) and Sp($N$) SYM. Although their full 1/16‑BPS indices coincide, we find discrepancies in their BMN‑sector indices. As the BMN indices restricted to the graviton sector are expected to be the same, this mismatch allows us to identify non‑graviton cohomologies. We explicitly find one of them in the SO(7) theory that is responsible for the mismatch of the BMN index. We also show that indices restricted to other one-loop cohomology truncations, in general, do not match under S-duality. If the conjecture of exactness of one-loop cohomology is correct, this suggests the presence of new  cohomology sub-rings  that match under S-duality with the letter-based truncations of the one-loop cohomology. This offers a way to check the one-loop exactness conjecture. 
}

\begin{document} 

\maketitle


\section{Introduction}

Supersymmetric black holes in AdS$_5$ have been extensively explored in type IIB supergravity \cite{Gutowski:2004ez, Gutowski:2004yv, Cvetic:2004ny, Chong:2005da, Chong:2005hr, Kunduri:2006ek}. Recent advances in their microscopic understanding demonstrate that the superconformal index of $\mathcal{N}=4$ SYM theory precisely reproduces their Bekenstein-Hawking entropy \cite{Hosseini:2017mds, Cabo-Bizet:2018ehj, Choi:2018hmj, Benini:2018ywd}. Beyond these solutions, it has been proposed that AdS black holes admit “hairy” generalizations - most notably the supersymmetric grey galaxy and the dual dressed black hole (DDBH) \cite{Kim:2023sig, Bajaj:2024utv, Choi:2024xnv, Choi:2025lck}. In the grey galaxy scenario, a BPS black hole coexists with a surrounding supersymmetric graviton gas, whereas the DDBH features a BPS black hole enveloped by a distant dual giant graviton. Evidence for these novel configurations also emerges from superconformal index calculations, further supporting the existence of hairy black hole states in AdS$_5$ \cite{Choi:2025lck}.

To probe the quantum structure of these black holes, it is natural to examine their constituent microstates. Thanks to the AdS/CFT correspondence, one can understand supersymmetric black hole microstates by studying dual BPS operators in \(\mathcal{N}=4\) SYM theory \cite{Maldacena:1997re}.  The systematic study of these operators via \(\mathcal{Q}\)-cohomology began with  \cite{Kinney:2005ej}, and subsequent works have classified the resulting spectrum into graviton and non-graviton (often loosely termed “black hole”) -- states that do not correspond to free gravitons in the bulk --  type \cite{Berkooz:2006wc, Grant:2008sk, Chang:2013fba}.

Recently, efforts to identify non-graviton (black hole) cohomologies have gained momentum. Beginning with \cite{Chang:2022mjp, Choi:2022caq}, there have been studies to find non-graviton cohomologies in \(\mathcal{N}=4\) SYM \cite{Choi:2023znd, Chang:2023zqk, Budzik:2023vtr, Choi:2023vdm, deMelloKoch:2024pcs} and this idea was extended to other holographic systems, including supersymmetric SYK models \cite{Chang:2024lxt} and the \(D1\)-\(D5\) system \cite{Chang:2025rqy}. A generalized notion of non-graviton states, termed fortuitous states, has also been proposed \cite{Chang:2024zqi}.\footnote{Accordingly, we will occasionally use fortuitous states to refer to non-graviton states.} 
Moreover, field theory evidence for the existence of supersymmetric grey galaxies was provided in \cite{Choi:2023znd, Choi:2023vdm, deMelloKoch:2024pcs}, where `hairy black hole' cohomologies were constructed by taking the product of high angular momentum gravitons with a seed non-graviton cohomology. Similarly, the existence of supersymmetric DDBHs was suggested by constructing analogous cohomologies in \cite{deMelloKoch:2024pcs}.  
Despite these advances, the detailed structure of the BPS black hole spectrum remains poorly understood perhaps due to its chaotic nature.
In this work, we take a small step toward addressing this problem by studying how the superconformal index of \(\mathcal{N}=4\) SYM organizes BPS black hole microstates. 

Throughout this paper, we will make use of the 1-loop non-renormalization conjecture. We precisely phrase and discuss this conjecture in section \ref{rev}. It roughly asserts that the 1-loop BPS spectrum remains unchanged even at strong coupling. Originally proposed in \cite{Grant:2008sk} and further supported in \cite{Chang:2022mjp}, this conjecture allows us to work entirely in the weakly coupled regime. Because the 1-loop sector is inherently weakly coupled, it is far more tractable than probing black hole microstates at strong coupling.

Moreover, a “bonus” U(1)  symmetry emerges at 1-loop \cite{Chang:2013fba}, allowing us to refine the index with a fugacity conjugate to this charge. This extra symmetry also admits several consistent one-loop truncations of \(\mathcal{N}=4\) SYM, most notably the BMN truncation, which retains three scalars, three fermions, and one holomorphic component of the field strength -- exactly matching the field content of the BMN matrix model Lagrangian \cite{Berenstein:2002jq, Kim:2003rza}. Although much simpler than the full theory, the BMN sector remains highly nontrivial, supporting numerous non-graviton cohomologies with entropy scaling as \(N^2\) \cite{Choi:2023vdm, Chang:2024lkw}.

A key feature of the BMN sector is that its single-letter index is a finite series in the fugacities. Consequently, the full BMN index -- expressed as a unitary matrix integral -- has only finitely many poles and can be evaluated exactly as a rational function in the fugacities. This form enables a systematic study of its structure, and in particular the denominator’s pattern reveals a recurring “tower” structure in the spectrum. Specifically, if the index contains a term of the form  
\begin{align}  
    I \supset \frac{1}{1-t^{j}} = 1 + t^{j} + t^{2j} + \cdots,  
\end{align}  
it implies the presence of a bosonic tower generated by a single-trace operator with charge $j$.  

In the first part of this paper, we propose a systematic method to find poles for computing the BMN index for SU($N$) gauge theories and analyze the general structure of its denominator. Our results reveal that the denominator structure of the BMN index is far richer than what can be explained by graviton cohomologies alone, hinting at a more intricate organization of black hole cohomologies in SU($N$).  
In particular, we show that the denominator contains towers of single-trace operators with charges \( j \) exceeding those of any single-trace graviton operators, indicating the presence of non-graviton cohomologies. (See \S \ref{gravtower} for a more detailed discussion.)  

In the second part, we examine \(\mathcal{Q}\)-cohomologies in SO($2N+1$) and Sp($N$) gauge theories. Since these theories are related by \(S\)-duality, their full \(1/16\)-BPS indices coincide. It is also shown that their 1-loop graviton cohomologies map to each other under the duality. However, we find that their BMN sectors do not generally match -- an explicit mismatch we confirm by computing the BMN indices for SO(7) and Sp(3).\footnote{SO(3)/Sp(1) and SO(5)/Sp(2) share identical Lie algebras and therefore always yield the same index. In particular, the integral computing the BMN index in one can be transformed to that of the other by change of integration variables.} Any discrepancy between the two BMN indices must arise from non-graviton states. In charge sectors where the indices differ, non-graviton operators appear in the BMN sector of one theory, while their \(S\)-dual counterparts lie outside the BMN sector of the other (and vice versa). Building on this observation, we outline a method to identify pairs of non-graviton cohomologies in SO($2N+1$) and Sp($N$) that are \(S\)-dual to each other, and we explicitly locate such a cohomology in the SO(7) theory. We also compare their bosonic “tower” structures and show that an operator generating a tower in one theory maps to a non-BMN operator in the dual theory. 

We also examine other 1-loop truncations of $\mathcal{N}=4$ SYM that corresponds gauge invariant operators that are built using a subset of all the BPS letters. We call such truncations ``letter-based'' truncations as they are defined by a subset of letters that is closed under $\mathcal{Q}$-action as discussed in detail in section \ref{letterbased}. Algebraically, the $\mathcal{Q}$-cohomology  forms a ring and the letter-based truncations define its sub-ring.
We find that the indices of SO(7) and Sp(3) ${\cal N}=4$  SYM over the other truncations also do not match.  These discrepancies signal the presence of non-graviton operators. 
The mismatches in 1-loop indices between two S-dual theories can serve as a diagnostic tool for detecting the presence of non-graviton operators in a given truncation, and provides a novel window into the \(S\)-duality properties of \(\mathcal{N}=4\) SYM. 
If we assume the 1-loop non-renormalization conjecture then the mismatch of operators in the BMN truncation and other letter-based truncations under S-duality has a tantalizing consequence: for every letter-based truncation sub-ring ${\cal T}$ there must be another sub-ring $\tilde {\cal T}$ that is isomorphic to ${\cal T}$ under S-duality. Existence of these such non-letter-based closed structures in 1-loop cohomology is then a non-trivial check of the 1-loop non-renormalization conjecture.

This note is organized as follows. In Section \ref{rev}, we review $\mathcal{N}=4$ SYM and discuss the classification of graviton and non-graviton cohomologies. We also introduce the notion of 1-loop truncation. In Section \ref{bmntr}, we develop a systematic method for computing the SU($N$) BMN index. Based on this, we analyze its structure and identify a ``non-graviton tower" that reveals features of the non-graviton spectrum. In Section \ref{sospbmn}, we study the SO($2N+1$) and Sp($N$) versions of $\mathcal{N}=4$ SYM, which are related by $S$-duality. In particular, we find a non-graviton operator in SO(7), inferred from the difference between the BMN indices of SO(7) and Sp(3). We further explore other 1-loop truncation sectors, showing that the mismatch between SO(7) and Sp(3) indices over those sectors indicates the presence of non-graviton cohomologies. We conclude with a discussion in Section \ref{discussion}.



\section{Review of $\mathcal{N}=4$ SYM}\label{rev}

In this section, we provide a brief review of \( \mathcal{N}=4 \) SYM and discuss the classification of BPS operators into graviton and non-graviton cohomologies.

The $\mathcal{N}=4$ SYM theory is a superconformal field theory. Its superconformal symmetry group is $\mathrm{PSU}(2,2|4)$, with a bosonic subgroup comprising of the 4-dimensional conformal group $\mathrm{SO}(4,2)$ and the R-symmetry group $\mathrm{SU}(4)$.  
This symmetry group includes 32 supercharges, 16 $\mathcal{Q}$s and 16 $\mathcal{Q}^{\dagger}$s\footnote{The supercharges ${\mathcal{Q}}^\dagger$ are Hermitian conjugate of supercharges $\mathcal{Q}$ and are sometimes denoted as $S$ and called conformal supercharges.}. 
Representations are characterized by the scaling dimension $E$, the Dynkin labels $(J_1, J_2)$ for the SO(4) global symmetry, and $(Q_1, Q_2, Q_3)$ for the R-symmetry group $\mathrm{SO}(6)\simeq \mathrm{SU}(4)$.  
Table \ref{charges-fields} lists all the fields that constitute the $\mathcal{N}=4$ SYM.  

\begin{table}[t]
\begin{center}
\begin{tabular}{c|c|c|c}
	\hline
    field & $E$ & ($Q_1$, $Q_2$, $Q_3$)  & ($J_1$, $J_2$) \\
	\hline $\phi_m$ & $1$ & ($-1,0,0$), ($0,-1,0$), ($0,0,-1$) & ($0,0$)\\
    $\bar\phi^m$ & $1$ & ($1,0,0$), ($0,1,0$), ($0,0,1$) & ($0,0$)\\
    $\psi_{m\pm}$ & $\frac{3}{2}$ & ($-\frac{1}{2},\frac{1}{2},\frac{1}{2}$),
    ($\frac{1}{2},-\frac{1}{2},\frac{1}{2}$), ($\frac{1}{2},\frac{1}{2},-\frac{1}{2}$)&
    ($\pm\frac{1}{2},\pm\frac{1}{2}$)\\
    $\bar\psi^m_{\dot\pm}$ & $\frac{3}{2}$ & ($\frac{1}{2},-\frac{1}{2},-\frac{1}{2}$),
    ($-\frac{1}{2},\frac{1}{2},-\frac{1}{2}$), ($-\frac{1}{2},-\frac{1}{2},\frac{1}{2}$)&
    ($\pm\frac{1}{2},\mp\frac{1}{2}$)\\
    \hline
    $A_{+\dot\pm}$ & $1$ & $(0,0,0)$ & $(1,0)$, $(0,1)$\\
    $A_{-\dot\pm}$ & $1$ & $(0,0,0)$ & $(-1,0)$, $(0,-1)$\\
    $f_{++},f_{+-},f_{--}$ & $2$ & ($0,0,0$) & ($1,1$), ($0,0$), ($-1,-1$)\\
    $f_{\dot{+}\dot{+}},f_{\dot{+}\dot{-}},f_{\dot{-}\dot{-}}$ & 
    $2$ & ($0,0,0$) & ($1,-1$), ($0,0$), ($-1,1$)\\
    $\lambda_\pm$ & $\frac{3}{2}$ & $(-\frac{1}{2},-\frac{1}{2},-\frac{1}{2})$ & 
    ($\pm\frac{1}{2},\pm\frac{1}{2}$)\\
    $\bar{\lambda}_{\dot\pm}$ & $\frac{3}{2}$ & $(\frac{1}{2},\frac{1}{2},\frac{1}{2})$ & 
    ($\pm\frac{1}{2},\mp\frac{1}{2}$)\\
    \hline
    $\partial_{+\dot\alpha}$ & $1$ &($0,0,0$) & ($1,0$), ($0,1$)\\
    $\partial_{-\dot\alpha}$ & $1$ &($0,0,0$) & ($-1,0$), ($0,-1$)\\
    \hline
\end{tabular}
\end{center}
\caption{The charges of elementary fields. Charges in the parenthesis 
are listed in the order of $m=1,2,3$ or $\dot\alpha=\dot{+},\dot{-}$ or in the order 
of the fields listed in the first column.}\label{charges-fields}
\end{table}

$1/16$-BPS operators are defined as those annihilated by one of sixteen supercharges \(\mathcal{Q}\) and its Hermitian conjugate, \(\mathcal{Q}^\dagger\). This condition is equivalent to,  
\begin{align}
    (\mathcal{Q}\mathcal{Q}^\dagger + \mathcal{Q}^\dagger \mathcal{Q})|O\rangle = 0 \quad \leftrightarrow \quad \mathcal{Q}|O\rangle = 0, \quad \mathcal{Q}^\dagger|O\rangle = 0.
\end{align}
Without losing generality, we choose $\mathcal{Q}\equiv \mathcal{Q}^4_{-}$, so that
\begin{align}
    2\{\mathcal{Q},\mathcal{Q}^{\dagger}\}\equiv H = E-J_1-J_2-Q_1-Q_2-Q_3.
\end{align}
The right hand side is positive semi-definite. The 1/16 BPS states saturate the BPS bound $E=J_1+J_2+Q_1+Q_2+Q_3$. 

At $g_{\text{YM}}=0$, the BPS fields that satisfy the bound are given as
\begin{equation}\label{BPS-fields}
  \bar\phi^m\ ,\ \ \psi_{m+}\ ,\ \ f_{++}\ ,\ \ 
  \bar\lambda_{\dot\alpha}\ ,\ \ \partial_{+\dot\alpha}\ ,
\end{equation}
which can be found from Table \ref{charges-fields} using their listed charges.
With these `letters', we construct gauge invariant operators. These are the BPS operators in the free theory.
However, most of these operators acquire anomalous dimensions already at 1-loop and only a subset remains protected. Identifying the general BPS operators is a challenging task. This problem can be reformulated in terms of \(\mathcal{Q}\)-cohomology \cite{Kinney:2005ej}, since BPS operators are in one-to-one correspondence with \(\mathcal{Q}\)-cohomology classes, as we elaborate below.
$\mathcal{Q}$ is a nilpotent operator so that one can interpret \(\mathcal{Q}\) as an exterior derivative and gauge-invariant operators as differential forms. Then \(\mathcal{Q}\mathcal{Q}^\dagger + \mathcal{Q}^\dagger \mathcal{Q}=H\) can be viewed as a harmonic operator, and BPS operators can be thought of as \(\mathcal{Q}\)-harmonic forms. Since harmonic forms are in one-to-one correspondence with cohomology classes, a BPS operator corresponds to a \(\mathcal{Q}\)-cohomology class.
The advantage of formulating the space of BPS operator as $\mathcal{Q}$-cohomology is that, as $\mathcal{Q}$-cohomology, they form a ring i.e. product of $\mathcal{Q}$-cohomology class is a $\mathcal{Q}$-cohomology class. Note that this closure under product does not necessarily hold for the harmonic representatives.

It has been conjectured for $\mathcal{N}=4$ SYM that the spectrum of operators that remains BPS at 1-loop is non-renormalized even in the strong coupling regime \cite{Kinney:2005ej,Janik:2007pm, Grant:2008sk}. Let us make this conjecture precise. Let the vector space of BPS operators at coupling $g$ be denoted by $V_g$. As remarked earlier, thinking of this space as $\mathcal{Q}$-cohomology allows us to multiply the elements endowing it with a further ring structure \footnote{The underlying vector space structure along with multiplication actually make it an algebra but we will only focus on the ring structure in this paper}. The five Cartan charges ${\vec q}\equiv (Q_1, Q_2, Q_3, J_1, J_2)$ commute with $H$ so we can simultaneously diagonalize them. We denote the eigenspaces of ${\vec q}$ as $V_g^{(\vec q)}$. %
Now we make two observations. For $g_1,g_2 \neq 0$,
\begin{itemize}
    \item there exists a vector space isomorphism between $V_{g_1}$ and $V_{g_2}$ that preserves ${\vec q}$. 
    \item there exists, in addition, a ring isomorphism between $V_{g_1}$ and $V_{g_2}$ that preserves ${\vec q}$.
\end{itemize}
The second observation is clearly stronger than the first. The first observation simply states that ${\rm dim} V_{g_1}^{(\vec q)}={\rm dim} V_{g_2}^{(\vec q)}$. This is expected because the dimension of the subspace is an integer and it is not expected to jump discontinuously as we vary the coupling continuously \cite{Chang:2013fba}. The discontinuous jumps in the protected spectrum would give rise to non-analyticity in the thermal partition function but for a gauge theory on the sphere with finte dimensional gauge group,   we don't expect the partition function to have any non-analyticity. 

The second observation is slightly more involved. As argued in \cite{Witten:1982df,Chang:2022mjp}, as the coupling is changed from one non-zero value to  another, the supercharge $\mathcal{Q}$ undergoes a similarity transformation $\mathcal{Q}_{g_1} \to \mathcal{Q}_{g_2} =M \mathcal{Q}_{g_1} M^{-1}$ with an invertible matrix $M$. The conjugate supercharge $\mathcal{Q}^\dagger$ undergoes similarity transformation by $M^\dagger$ and because $M$ is not necessarily unitary, $H$ undergoes a non-trivial transformation. Since we only use $\mathcal{Q}$ to definite cohomology, $M$ yields the explicit isomorphism  between them. 
As charges are discrete, we expect $M$ to commute with ${\vec q}$. 
We formulate 
the 1-loop renormalization conjecture which promote both of the above observations to 1-loop. 

In order to state the conjectures, let us first define the 1-loop  $\mathcal{Q}$-cohomology $V_{1\text{-loop}}$.  The action of $\mathcal{Q}$ at this order is directly obtained from the classical interacting Lagrangian of ${\cal N}=4$.
\begin{eqnarray}\label{half-loop-Q}
  &&[\mathcal{Q},\bar\phi^m]=0\ ,\ \{\mathcal{Q},\bar\lambda_{\dot\alpha}\}=0\ ,\ 
  \{\mathcal{Q},\psi_{m+}\}=-ig_{\rm YM}\epsilon_{mnp}[\bar\phi^n,\bar\phi^p]\\
  &&[\mathcal{Q},f_{++}]=-ig_{\rm YM}[\psi_{m+},\bar\phi^m]\ ,\ 
  [\mathcal{Q},D_{+\dot\alpha}](\cdots)=-ig_{\rm YM}[\bar\lambda_{\dot\alpha},(\cdots)\}\ .
  \nonumber
\end{eqnarray}
Due to the appearance of $g_{YM}$ on the right hand side, this the action of $\mathcal{Q}$ is sometimes referred to as being at half-loop. Both $\mathcal{Q}$ and $\mathcal{Q}^\dagger$ are ${\cal O}(g_{YM})$ and their anti-commutator $\mathcal{Q}\mathcal{Q}^\dagger+\mathcal{Q}^\dagger \mathcal{Q}$ is ${\cal O}(g_{YM}^2)$ as expected of a 1-loop Hamiltonian. Now we are ready to state the 1-loop non-renormalization conjecture. 
\begin{itemize}
    \item {\bf Conjecture 1}: There exists a vector space isomorphism between $V_{g}$ and $V_{1\text{-loop}}$ that preserves ${\vec q}$.
\item {\bf Conjecture 2}: There exists, in addition, a ring isomorphism between $V_{g}$ and $V_{1\text{-loop}}$ that preserves ${\vec q}$.
\end{itemize}
In this paper we will provide a way to verify the conjecture $1$ and conjecture $2$. 


\subsection{BPS partition function and the Superconformal index}
In order to discuss the 1-loop cohomology, it is useful to define the partition function over BPS states.
\begin{align}
    Z_{\rm BPS}=\text{Tr}_{V} \,\,p^{E-J_2} q^{E-J_1} a^{-Q_2-Q_3} b^{-Q_1-Q_3} c^{-Q_1-Q_2},
\end{align}
Here we introduced fugacities for a convenient set of  spanning vectors in the five dimensional charge space. The trace is taken over the BPS subspace i.e. the space defined by the relation $E=J_1+J_2+Q_1+Q_2+Q_3$. The advantage of picking the fugacities for this particular set of spanning charges is that the superconformal index -- in which only the $\mathcal{Q}$-commuting charges 
are counted -- is obtained from $Z_{\rm BPS}$ by setting $abc=pq$ and inserting an extra $(-1)^F$ within the trace. The relation $abc = pq$ is often solved with the substitution
\begin{align}
    p=t^3 y, \quad q=t^3/y, \quad a=t^2 v, \quad b = t^2 u , \quad c= t^2/(u v). 
\end{align}
To simplify calculations, we will often make the above replacement with $y=u=v=1$ and just keep track of the charge with respect to the variable $t$.

The action of 1-loop Hamiltonian on the free BPS operators preserves the number of letters within the operator. This symmetry is known as the bonus $U(1)_Y$ symmetry. 
The 1-loop index can then be further refined by this symmetry. We discuss this bonus symmetry and the associated refinement of the index in section \ref{letterbased}.

\subsection{Graviton and non-graviton cohomologies}\label{coh}
Now we turn to the classification of 1-loop BPS cohomologies into two categories: graviton cohomologies and non-graviton cohomologies.
We sometimes abuse the terminology `black hole cohomology' to refer to non-graviton cohomologies, even though they may not be strictly equivalent.\footnote{By definition, black hole microstates are contained within the non-graviton cohomologies. However, it may not be always the case that the entirety of non-graviton cohomologies corresponds to black hole cohomologies.}

Gravitons are defined as multi-graviton states, which are products of single graviton states.
Single graviton states correspond to the irreducible representations of single-particle (single-trace) states appearing in supergravity in \(AdS_5 \times S^5\). 
The single-trace graviton spectrum consists of \(\frac{1}{2}\)-BPS chiral primaries, denoted as \(|n\rangle\), along with their superconformal descendants in PSU$(2,1|3)$ (See Table \ref{graviton-multiplet} for the state contents). 
This multiplet where $|n\rangle$ is a superconformal primary single graviton state is called $S_n$ multiplet. The primary state is explicitly given by
\begin{align}\label{Sn primary}
    |n\rangle= \text{Tr}\left[\bar{\phi}^{(i_1}\cdots\bar{\phi}^{i_n)}\right].
\end{align}
\begin{table}[t]
\begin{center}
\begin{tabular}{c|c|c|c|c}
	\hline
    $(-1)^F E^\prime$ & $J^\prime$ & $R_1^\prime$ & $R_2^\prime$ & construction\\
	\hline $n$ & $0$ & $n$ & $0$ & $|n\rangle$ \\
    $-(n+\frac{1}{2})$ & $\frac{1}{2}$ & $n-1$ & $0$ &
    $\overline{\mathcal{Q}}_{m\dot\alpha}|n\rangle$ \\
    $n+1$ & $0$ & $n-2$ & $0$
    & $\overline{\mathcal{Q}}_{m\dot{+}}\overline{\mathcal{Q}}_{n\dot{-}}|n\rangle$ \\
	$-(n+1)$ & $0$ & $n-1$ & $1$ &
    $\mathcal{Q}^m_+|n\rangle$\\
    $n+\frac{3}{2}$ & $\frac{1}{2}$ & $n-2$ & $1$ &
    $\mathcal{Q}^m_+\overline{\mathcal{Q}}_{n\dot\alpha}|n\rangle$ \\
    $-(n+2)$ & $0$ & $n-3$ & $1$ &
    $\mathcal{Q}^m_+\overline{\mathcal{Q}}_{n\dot{+}}\overline{\mathcal{Q}}_{p\dot{-}}|n\rangle$ \\
    $n+2$ & $0$ & $n-1$ & $0$ & $\mathcal{Q}^m_+\mathcal{Q}^n_+|n\rangle $ \\
    $-(n+\frac{5}{2})$ & $\frac{1}{2}$ & $n-2$ & $0$ &
    $\mathcal{Q}^m_+\mathcal{Q}^n_+\overline{\mathcal{Q}}_{p\dot\alpha}|n\rangle$ \\
    $n+3$ & $0$ & $n-3$ & $0$ &
    $\mathcal{Q}^m_+\mathcal{Q}^n_+\overline{\mathcal{Q}}_{p\dot{+}}\overline{\mathcal{Q}}_{q\dot{-}}|n\rangle$ \\
    \hline
\end{tabular}
\caption{The state contents of the PSU($1,2|3$) multiplet $S_n$. 
For low $n$'s, the rows with negative $R_1^\prime$ are absent. $|n\rangle$ schematically 
denotes the superconformal primaries.}\label{graviton-multiplet}
\end{center}
\end{table}
In the strict large N limit ($N\to \infty$), the \(S_n\) multiplets are independent of one another, as no trace relations exist. 
Furthermore, all BPS states are generated by \(S_n\) multiplets. However, at finite N, trace relations reduce the number of graviton states, as \(S_n\) operators with \(n > N\) can be written as sums of products of operators with \(n \leq N\) .
For example, in SU(2) theory, it suffices to consider $S_2$ multiplet; states in $S_{n\geq 3}$ multiplets are all expressed as products of states in $S_2$ multiplet.

The trace relations not only reduce the number of BPS states but also give rise to new types of BPS states.  
For instance, an operator $O_N$ in SU($N$) theory can become $\mathcal{Q}$-closed as a result of the trace relation specific to SU($N$), such that:  
\begin{align}  
    \mathcal{Q} O_N = (\text{trace relation in  SU($N$)})=0.  
\end{align}  
Since the trace relation in SU($N$) is absent in gauge theories 
For higher ranks $N' > N$, the operator $O_N$ remains $\mathcal{Q}$-closed in SU($N$) and in gauge groups of lower rank. Moreover, if $O_N$ is not $\mathcal{Q}$-exact, it represents a nontrivial element of the $\mathcal{Q}$-cohomology that is not present in arbitrary theories. 
Such states are referred to as non-graviton states. They are also commonly referred to as fortuitous states because a given fortuitous state is in the cohomology only for particular gauge group(s). \cite{Chang:2024zqi}. 

To find non-graviton states that come from trace relations, the most commonly used strategy is to use the superconformal index \cite{Chang:2022mjp,Choi:2023znd,Choi:2023vdm,deMelloKoch:2024pcs,Chang:2025rqy}, which goes as follows.
First, compute the the `graviton index'.
It involves listing all the independent graviton operators and counting them, accounting for signs arising from fermions (see subsection  \ref{gravind} for the explanation and an example).
Next, subtract the graviton index from the full index to obtain the `non-graviton index' whose contribution comes solely from non-graviton cohomologies.
Finally, find an explicit representative of the non-graviton cohomology using the charges that are read off from the non-graviton index.

\subsection{Graviton Index}\label{gravind}
In this subsection, we review the computation of the  graviton index using the method outlined in \cite{Choi:2023znd}.
We begin by considering the chiral primary states \( |n\rangle \). Since all scalar fields are symmetrized within the trace (see \eqref{Sn primary}), they can be treated as diagonal matrices, allowing us to regard them as mutually commuting variables. In SU($N$), each scalar field contributes \( N-1 \) independent eigenvalues. The chiral primary generators \( |n\rangle \) are then expressed as Weyl-invariant polynomials constructed from these eigenvalues.  

Next, we consider the superconformal descendants of the chiral primaries in the free theory limit. In this limit, supersymmetry transformations of diagonal fields only involve the diagonal components of their superpartner fields. As a result, we can restrict all elementary fields to their diagonal components.  
The polynomials representing the chiral primaries and their superconformal descendants in \( S_{n \leq N} \) are referred to as generators. To obtain the graviton index, we enumerate all independent polynomials that can be formed from these generators and construct a (graded) partition function.

\subsubsection{Example: SU(2) Hall-Littlewood graviton index}\label{hall-little}
As the simplest example, we compute the graviton index in the Hall-Littlewood sector of SU(2).  
The Hall-Littlewood sector belongs to a $1/4$-BPS sector, consisting of two scalars $\bar{\phi}^1, \bar{\phi}^2$ and one fermion $\psi_{3+}$.

When considering graviton operators, all fields can be treated as diagonal matrices. Therefore, we only consider Cartan generators. In SU(2), the  fields take the form:  
\begin{align}
    \bar{\phi}^1=\begin{pmatrix} x &0\\0 &-x\end{pmatrix},\quad
    \bar{\phi}^2=\begin{pmatrix} y &0\\0 &-y\end{pmatrix}
,\quad
    \psi_{3+}=\begin{pmatrix} \psi_3 &0\\0 &-\psi_3\end{pmatrix}.
\end{align}
For SU(2), the \( S_2 \) multiplet generates all the multi-graviton states.  
The states in the \( S_2 \) multiplet within the Hall-Littlewood sector are given by:
\begin{align}
    |2\rangle &= \text{Tr}\left[\bar{\phi}^1\bar{\phi}^1\right],~~\text{Tr}\left[\bar{\phi}^1\bar{\phi}^2\right],~~\text{Tr}\left[\bar{\phi}^2\bar{\phi}^2\right],\nonumber\\
    \bar{\mathcal{Q}}^3_+ |2\rangle &= \text{Tr}\left[\bar{\phi}^1\psi_{3+}\right], ~~\text{Tr}\left[\bar{\phi}^2\psi_{3+}\right].
\end{align}
Thus, the independent generators are:  
\begin{align}\label{su2gens}
    x^2, \quad y^2, \quad xy, \quad x\psi_3, \quad y\psi_3.
\end{align}
The SU(2) multi-graviton states can be thought of as monomials that are generated by the generators above.\footnote{For higher rank, the computation is more involved. For example, in SU(3), the fields take the form 
\begin{align}
    \bar{\phi}^1 = \text{diag}(x_1,x_2,-x_1-x_2),\quad \bar{\phi}^2 = \text{diag}(y_1,y_2,-y_1-y_2),\quad \psi_{3+} = \text{diag}(\psi_{31},\psi_{32},-\psi_{31}-\psi_{32}),
\end{align}
and the generators are given by the following polynomials.
\begin{align}
    x_1^2+x_2^2+(x_1+x_2)^2, y_1^2+y_2^2+(y_1+y_2)^2, x_1y_1+x_2y_2+(x_1+x_2)(y_1+y_2),\nonumber\\
x_1\psi_{31}+x_2\psi_{32}+(x_1+x_2)(\psi_{31}+\psi_{32}), y_1\psi_{31}+y_2\psi_{32}+(y_1+y_2)(\psi_{31}+\psi_{32})
\end{align}
}
We count multigraviton states by classifying them into bosonic and fermionic states.
\begin{itemize}
    \item Bosonic Sector ($\psi^0$):
    The bosonic sector consists of monomials of the form $x^n y^m$, with $n+m$ even.  
These are all the monomials that are invariant under $(x,y) \to (-x,-y)$, leading to the generating function:
\begin{align}
    \frac{1}{2} \left(\frac{1}{(1-x)(1-y)}+\frac{1}{(1+x)(1+y)}\right) = \frac{1+xy}{(1-x^2)(1-y^2)}.
\end{align}
    \item Fermionic Sector ($\psi^1$):
    The fermionic sector consists of monomials of the form $x^n y^m \psi$, with $n+m$ odd.  
The generating function is:
\begin{align}
    \psi_3 \frac{1}{2} \left(\frac{1}{(1-x)(1-y)} - \frac{1}{(1+x)(1+y)}\right) = \frac{\psi_3 (x+y)}{(1-x^2)(1-y^2)}.
\end{align}
\end{itemize}
In the superconformal index, the letters $\bar \phi_1, \bar \phi_1$ and $\psi_3$ contribute the fugacities $a,b$ and $ab$ respectively.
Hence the graviton index is obtained by substituting $x=a, y=b, \psi_3 = ab$, and adding the two generating functions obtained above with a negative sign for the fermionic states.
\begin{align}\label{su2-grav-ind}
    I = \frac{1 + ab - (ab)(a+b)}{(1-a^2)(1-b^2)}.
\end{align}
In fact, this sector does not host any non-graviton cohomologies.  
Thus, the graviton index in this sector should match the full Hall-Littlewood index, which is indeed the case.

\subsection{Letter-based truncation}\label{letterbased}

In this subsection we introduce the notion of a letter-based 1-loop truncation and classify them in $\mathcal{N}=4$ SYM. 
Under the 1‑loop renormalization conjecture, we may restrict our analysis to the 1‑loop \(\mathcal{Q}\)‑cohomology rather than the full‑loop problem, simplifying computations significantly. Notably, Chang and Yin \cite{Chang:2013fba} discovered an emergent U(1) “bonus” symmetry at 1-loop whose charge \(Y\) simply counts the number of letters in each operator.  The 1-loop supercharge \(\mathcal{Q}\), and $\mathcal{Q}^\dagger$ carry the charge \(Y=1\) and $Y=-1$ respectively. As a result the 1-loop Hamiltonian is neutral under $Y$ and that means that the sectors with different \(Y\) never mix. The action of 1-loop $\mathcal{Q}$ however increase the $Y$ and hence the number of letters by $1$.

Exploiting this symmetry, one defines a refined 1‑loop index in \(\mathcal{N}=4\) SYM as 
\begin{align}\label{1loopind}
    I_{1\text{-loop}}&=\text{Tr}(-1)^Fp^{E-J_2-Y} q^{E-J_1-Y} a^{Y-Q_2-Q_3} b^{Y-Q_1-Q_3} c^{Y-Q_1-Q_2}.\notag\\
    &=\text{Tr}(-1)^Fp^{E-J_2} q^{E-J_1} a^{Q_2-Q_3} b^{Q_1-Q_3} c^{Q_1-Q_2} \Big(\frac{abc}{pq}\Big)^Y.
\end{align}
The single-letter index of the 1-loop truncation is given by  
\begin{align}\label{truncind}
    f = 1 - \frac{(1 - a)(1 - b)(1 - c)}{(1 - p)(1 - q)}.
\end{align}
As is clear from the equation \eqref{1loopind}, setting $abc=pq$ restores $I_{1\text{-loop}}$ to the familiar superconformal index. 

Due to the bonus $U(1)_Y$ symmetry at 1-loop, the theory admits various “letter‑based” truncations at 1-loop, each consisting of a subset of BPS letters closed under the half‑loop \(\mathcal{Q}\) action.  Closure means that \(\mathcal{Q}\) acting on any letter in the subset produces only letters within that subset.  Let us define the truncation precisely. Let the full set of free BPS letters be 
\begin{align}
\{W_A\} = \{\bar\phi^m,\;\psi_{m+},\;f_{++},\;\bar\lambda_{\dot\alpha},\;\partial_{+\dot\alpha}\}
\end{align}
as in \eqref{BPS-fields}. A letter-based truncation is a the ring of gauge invariants formed with a subset of BPS letters \(\{W_a\}\subset\{W_A\}\) that obeys the following conditions.  
\begin{align}\label{cons-trunc}
    \mathcal{Q}W_a = f(W_a); 
\qquad\qquad 
\mathcal{Q}W_A = g(W_A, W_a),\quad {\rm with} \quad g(W_A, W_a)\bigl|_{W_A = 0} = 0.
\end{align}
These conditions mean that setting the complementary letters \(W_A\) to zero guarantees no new letters outside \(\{W_a\}\) are generated by the $\mathcal{Q}$ action.

To identify such subsets, it is useful to note that the 1-loop hamiltonian commutes with the $U(1)_Y$ symmetry. Hence working with a sector specified by setting any of the $5$ charges in equation \eqref{1loopind} to zero gives rise to a sector that is closed under multiplication. This corresponds to setting one or more of the fugacities $\{a,b,c,p,q\}$ to zero. Indeed, we identify two types maximal 1-loop truncations
\begin{align}
    &\{\bar\phi^i,\psi_{i+},f_{++},\bar\lambda_{\dot+},D_{+\dot+}\},\,\,\qquad f=1 - \frac{(1 - a)(1 - b)(1 - c)}{(1 - p)}.\\
    & \{\bar\phi^{1,2},\psi_{3+},\bar\lambda_{\dot\alpha},D_{+\dot\alpha}\},\qquad \qquad f=1 - \frac{(1 - a)(1 - b)}{(1 - p)(1 - q)}.
\end{align}
It can be checked directly from equation \eqref{half-loop-Q}, that the above subsets obey the necessary conditions \eqref{cons-trunc} to be consistent truncation. 
The other maximal truncations correspond to setting $p$, $a$ or $b$ to zero instead of $q$ or $c$. Even smaller truncations are obtained by setting more than one fugacity to $0$. 

In this paper, we focus on the BMN truncation, defined by the letters \(\{\bar\phi^m,\psi_{m+},f_{++}\}\), which corresponds to taking \(p\to0\) in the first truncation above. Note that both supersymmetric derivatives are absent from this truncation. This is  advantageous in computing the gauge theory index restricted over this sub-sector using residue integration. The next section, section \ref{bmntr}, is dedicated to the detailed study of the BMN truncation. The gauge theory index over the rest of the truncations is studied further in section  \ref{1ltrun}.




\section{BMN truncation and SU($N$) BMN index}\label{bmntr}

In this section, we investigate the BMN sector of SU($N$) \( \mathcal{N}=4 \) SYM. Our goal is to analyze the spectrum by studying the BMN index, with a particular focus on its structure arising from non-graviton cohomologies.  

It was shown in \cite{Kim:2003rza} that classical \( \mathcal{N}=4 \) SYM admits a consistent truncation to the BMN matrix model \cite{Berenstein:2002jq}. This means the classical equations of motion remain preserved within this sector, and are identical to those of the BMN matrix model.
The BMN matrix model is a mass-deformed version of the BFSS model \cite{Banks:1996vh}, providing a dual description of M-theory in a supersymmetric plane-wave background. The BMN action takes the form:
\begin{align}
    S &= \int dt \; \text{tr} \Bigg[ \frac{1}{2}(D_t X^I)^2 - i\theta D_t \theta + \frac{1}{4}[X^I, X^J]^2 + \theta \Gamma^I [X^I, \theta] \nonumber\\
    & \quad -\frac{1}{2} \left(\frac{m}{3}\right)^2(X^a)^2 - \frac{1}{2}\left(\frac{m}{6}\right)^2(X^i)^2 + \frac{m}{4}i\theta \Gamma^{123}\theta + \frac{m}{3}i\epsilon_{abc}X^a X^b X^c \Bigg],
\end{align}
where $I=1,\ldots,9$, $a=1,2,3$, and $i=1,\ldots,6$. Here, $D_t = \partial_t - i[\omega, \cdot]$ is the covariant derivative, with $\omega$ being the gauge field. The fields $X^I$ represent scalars, and $\theta$ represents fermions, with the system respecting SO(1,3) $\times$ SO(6) symmetry in the Lorentzian theory.
The correspondence between $\mathcal{N}=4$ SYM fields and BMN fields is as follows: the six scalars $X^I$ map to $\phi^m$, $\bar{\phi}^m$; the three scalars $X^a$ map to $A_a$; $\theta$ maps to $\psi_m$ and $\lambda_\alpha$; and $\omega$ maps to $A_0$.  

For 1-loop cohomology calculations involving the classical supercharge $\mathcal{Q}$, the BPS cohomology problem in SYM consistently truncates to the `BMN sector'. This follows from the fact that the BMN sector forms a closed algebra under the action of $\mathcal{Q}$ at the classical level, with the relations:
\begin{align}\label{Q-BMN}
    [\mathcal{Q}, \bar{\phi}^m] &= 0, \nonumber\\
    \{\mathcal{Q}, \psi_{m+}\} &= -i g_{\text{YM}} \epsilon_{mnp} [\bar{\phi}^n, \bar{\phi}^p], \\
    [\mathcal{Q}, f_{++}] &= -i g_{\text{YM}} [\psi_{m+}, \bar{\phi}^m],\nonumber
\end{align}
where the fields $\bar{\phi}^m$, $\psi_{m+}$, and $f_{++}$ are letters that consist the BMN sector.
While the consistent truncation in SYM does not extend to the quantum theory, it is conjectured that the 1-loop cohomology in $\mathcal{N}=4$ SYM remains non-renormalized. This suggests that the 1-loop cohomology spectrum found in the BMN sector forms a closed subsector i.e. a sub-ring within the full theory.

The superalgebra that acts within the BMN sector  reduces from PSU$(2,1|3)$, which is the commutant of a single supercharge, to PSU$(1|3)$. Note that the bosonic subalgebra is compact indicating the absence of derivative operators in the truncation.  Among the BPS single-graviton states in Table~\ref{graviton-multiplet}, only $|n\rangle$, $\bar{\mathcal{Q}}^m_{+}|n\rangle$, and $\bar{\mathcal{Q}}^m_{+} \bar{\mathcal{Q}}^n_{+}|n\rangle$ states are retained in this truncation.
\begin{align}\label{bmnsusn}
    |n\rangle=u_{n}^{(i_1\cdots i_n)} &= \text{Tr}\left[\bar{\phi}^{(i_1}\cdots\bar{\phi}^{i_n)}\right],\nonumber\\
    \bar{\mathcal{Q}}^m_{+}|n\rangle=v_{n}^{(i_1\cdots i_{n-1})}~_{j}&=\text{Tr}\left[\bar{\phi}^{(i_1}\cdots\bar{\phi}^{i_{n-1})}\psi_j\right]-(\text{trace}),\\
    \bar{\mathcal{Q}}^m_{+}\bar{\mathcal{Q}}^n_{+}|n\rangle= w_{n}^{(i_1\cdots i_{n-1})}&= \text{Tr}\left[\bar{\phi}^{(i_1}\cdots\bar{\phi}^{i_{n-1})}f+\frac{1}{2}\epsilon^{jk(i_p}\sum_{p=1}^{n-1}\bar{\phi}^{i_1}\cdots\bar{\phi}^{i_{p-1}}\psi_j\bar{\phi}^{\phi_{p+1}}\cdots\bar{\phi}^{i_{n-1})}\psi_k\right]\nonumber
\end{align}
The multiplicity of each of the above states in $S_n$ multiplets is
\begin{align}
    \frac{(n+1)(n+2)}{2},~~~ n(n+2), ~~~\frac{n(n+1)}{2}
\end{align}
respectively.

Now let us write down the BMN index of SU($N$) theory. Since the relevant letters are $\bar{\phi}^m$, $\psi_{m+}$, and $f_{++}$ the single particle index $f$ is given as
\begin{align}
    f=a+b+c-ab-bc-ca+abc.
\end{align}
Recall that $a,b,c$ are fugacities for the charges $Q_1+\frac{J_1+J_2}{2}$, $Q_2+\frac{J_1+J_2}{2}$, and $Q_3+\frac{J_1+J_2}{2}$, respectively.

The index of the $U(N)$ adjoint theory whose single letter index takes the form
\begin{align}\label{single-gen}
    f = \sum_{x \in P} x - \sum_{y \in Z} y,
\end{align}
where $x$'s and $y$'s are monic monomials in the fugacity,
is given by the $N$-variable integral,
\begin{align}\label{general-int}
    {\cal I}_N (x_1, \ldots, x_d) = \frac{1}{N!}\Big(\frac{\prod_{y \in Z}1-y }{\prod_{x\in P}1-x }\Big)^N\oint \prod_{i =1}^N \frac{dz_i}{2\pi i z_i} \prod_{i\neq j}(1-\frac{z_i}{z_j})  \frac{\prod_{y \in Z}1-y \frac{z_i}{z_i}}{\prod_{x\in P}1-x \frac{z_i}{z_i}}.
\end{align}
The familiar limits of the superconformal index of ${\cal N}=4$ super Yang-Mills correspond to the following.  
\begin{itemize}
    \item Half BPS index: $P=\{a\}$ and $Z=\emptyset$.
    \item Hall-Littlewood index: $P=\{a,b\}$ and $Z= \{ab\}$
    \item BMN index: $P=\{a,b,c,abc\}$ and $Z= \{ab,bc,ca\}$
\end{itemize}
As a result, the index restricted to the BMN sector {\emph a.k.a.} BMN index is
\begin{equation}\label{int0}
     I_{\mathrm{BMN}}=\frac{\kappa^{N-1}}{N!}\oint_{\substack{|z_i|=1}}\prod_{i=1}^{N}\frac{dz_i}{2\pi i z_i}\prod_{\substack{j=1\\j\neq i}}^{N}\frac{\Big(1-\frac{z_i}{z_j}\Big)\Big(1-ab\frac{z_i}{z_j}\Big)\Big(1-bc\frac{z_i}{z_j}\Big)\Big(1-ca\frac{z_i}{z_j}\Big)}{\Big(1-abc\frac{z_i}{z_j}\Big)\Big(1-a\frac{z_i}{z_j}\Big)\Big(1-b\frac{z_i}{z_j}\Big)\Big(1-c\frac{z_i}{z_j}\Big)}
\end{equation}
where $\kappa=\frac{(1-ab)(1-bc)(1-ca)}{(1-abc)(1-a)(1-b)(1-c)}$.
Also, We assume $|a|<1,|b|<1,|c|<1$.
Since the single-letter index for this sector is a finite series in fugacities, there are finite numbers of poles. Thus the index is expressed as a rational function. Existence of derivatives would have  infinitely many terms in the single-letter index and would give rise to infinitely many poles in the gauge theory index complicating the exact  computation substantially. We will discuss the exact computation of this integral using method of residues in section \ref{pole res}. For now, let us analyze this index in the large $N$ limit.

At large $N$ we change the integration variable from $z=e^{i\alpha}$'s to their normalized density $\rho(\alpha)$ on the unit circle. In these variables, the action of the matrix integral scales as $N^2$ which allows for its computation using saddle point method. The saddle point equation is
\begin{align}\label{saddle}
    \int d\alpha\rho(\alpha)V'(\alpha_i-\alpha)=0,
\end{align}
where $V$ is defined such that the integral we want to evaluate is given by \begin{align}
    \int \prod_{i=1}^N d\alpha_i\exp\left[-\sum_{i\neq j}V(\alpha_i-\alpha_j)\right].
\end{align}
The authors of \cite{Choi:2023vdm} find a deconfined saddle of the integral expression of the BMN index \eqref{int0}:  
\begin{align}\label{su saddle}
    \rho(\alpha) = \frac{3}{4\pi^3}(\pi^2 - \alpha^2), \quad -\pi < \alpha < \pi,
\end{align}
The corresponding free energy at large \( N \) scales as $O(N^2)$ and is given by  
\begin{align}
    \log I = -\frac{3N^2}{2\pi^2} \Delta_1 \Delta_2 \Delta_3.
\end{align}
where $\Delta_I$ are defined as $e^{-\Delta_1}= a$, $e^{-\Delta_2}= b$, $e^{-\Delta_3}= c$. 
The entropy can then be obtained via a Legendre transformation:  
\begin{align}
    S = \text{Ext} \left[ \log I(\Delta_I) + \sum_I (Q_I + J_L) \Delta_I \right] = 2\pi \sqrt{\frac{2(Q_1+J_L)(Q_2+J_L)(Q_3+J_L)}{3N^2}}.
\end{align}
It is important to note that the saddle point analysis in \cite{Choi:2023vdm} is valid only for small \( \Delta_I \), corresponding to the regime where charges are small: \( \frac{Q_I+J_L}{N^2} \ll 1\).
In this limit, one can compare the entropy of the BPS black hole in the full $1/16$ BPS sector with the same charges. The entropy is given as $S=2\pi \sqrt{\frac{2(Q_1+J_L)(Q_2+J_L)(Q_3+J_L)}{N^2}}$, which is $\sqrt{3}$ times larger than that of the `black hole' saddle in the BMN sector \cite{Choi:2021lbk}.

On the other hand, the behavior of the BMN index in the large charge limit differs significantly from that of the full \(\mathcal{N}=4\) SYM theory. Let us assume $j\equiv Q+J_L$, $Q_1=Q_2=Q_3=Q$ for simplicity. The large charge limit corresponds to $j\gg N^2$.
In the full sector, the entropy of large black holes scales as \( S_{\text{BH}} \approx 2\pi \left(\frac{N^2 j^2}{2}\right)^{1/3} \). However, the BMN index grows much more slowly. This slow growth can be attributed to the absence of derivatives in the sector, which limits the number of accessible states. For a related discussion see \cite{Choi:2023vdm}.

To quantify this suppression, we estimate an upper bound on the growth rate of the BMN index by considering the free BMN \textit{partition function}, which provides a natural upper limit on the number of states. The integral representation of the free BMN partition function is given as follows. For real and positive fugacities $a,b,c$ we have,
\begin{align}\label{partint}
	Z = \frac{\kappa^N}{N!}\oint \prod_{i=1}^N \frac{dz_i}{2\pi i z_i} \prod_{j\neq i}^N \frac{\big(z_i - z_j\big) \big(z_i + ab z_j\big) \big(z_i + bc z_j\big) \big(z_i + ca z_j\big)}{\big(z_i - abcz_j\big) \big(z_i - a z_j\big) \big(z_i - b z_j\big) \big(z_i - c z_j\big)}\nonumber\\
	<\left(\frac{2(1+ab)(1+bc)(1+ca)}{(1-abc)(1-a)(1-b)(1-c)}\right)^{N^2}.
\end{align}
Let $a=b=c=e^{-\beta}$.
At high temperature ($\beta \to 0+$), where the fugacities approach unity, the right hand side of \eqref{partint} behaves as $N^2\log(\frac{16}{3\beta^4})$.
From this, an entropy bound can be obtained as follows  $N^2\left(4-\log48+4\log\frac{j}{N^2}\right)$, leading to the conclusion that the entropy in this sector remains much smaller than that of large black holes in the full theory. 
This function scales as \( \sim N^2 \log j \), meaning that its growth rate is at most proportional to \( N^2/j \). This is already much slower than the black hole entropy formula, which scales as \( (N^2/j)^{1/3} \). 

Let us note that at strict large $N$ limit, the index is equivalent to the graviton index. Therefore, it can be obtained by summing over all the contributions from $S_{k}$ multiplets where $k\in \mathbb{N}$:
\begin{align}\label{SUinf}
    I_{U(\infty)}=\prod_{n=1}^{\infty}\frac{1}{(1-a^n)(1-b^n)(1-c^n)}
\end{align}
The indicial entropy $S(E)$ scales as $\sim E^{1/2}$ for large $E$.

\subsection{Intermission: Residue integration}\label{pole res}
This section is dedicated to computing the index integrals of the type \eqref{general-int} using the method of residues. To understand the structure of the multi-variable poles, we will first work with the general form of the integral \eqref{general-int} and then specialize to particular limits of the superconformal index and classify their poles. We will assume that all the monomials $x\in P$ and $y\in Z$ obey  $|x|, |y|<1$.  The elements of set $P$ and $Z$ give rise to the terms that cause poles and zeros of the integral respectively. 
Now we will discuss the classification of poles for such an integral. 

First off, it is convenient to shift the pole at $z_i=0$ to $z_i= u$, for some fixed $u$ by changing the $z_i$ in the denominator of the first term to $z_i - u$ i.e.
\begin{align}
    \frac{1}{2\pi i z_i}\to \frac{1}{2\pi i (z_i-u)}.
\end{align} 
Note that, despite explicit $u$ dependence of the integrand, the integral is independent of $u$. This can be seen by redefining $z_i\to u z_i$. In the rest of our discussion we will always work with this shift.

The multi-variable residue integral in \eqref{general-int} is done by first doing a residue integral in one variable, and then the next and so on. 
Unless otherwise explicitly stated, we will compute the residue in $z_1$ first, then $z_2$ and so on. 
We label the combined pole in \emph{all}  the variables by an ordered ``pole-tuple'' $(z_1^*,\ldots, z_N^*)$, where the order indicates the order of integration. In general, each of the $z_i^*$ in the pole-tuple is a function of $u$ and $x\in P$, in fact, as it turns out, it takes the form $z_i^* = u {\vec x}^{\vec v}$ for some $d$ dimensional vector $v$ with integer components. Here we have introduced the notation ${\vec x}^{\vec v}\equiv \prod_{x_a\in P} x_a^{v_a}$.

Given an ordered pole-tuple, we can associate a directed graph to it. The entries of the pole-tuple for the vertices of the graph. We draw an arrow from $z_i^*$ to $z_j^*$ if $z_j^* = x z_i^*$ for some $x\in P$. Naturally, this arrow is labeled by the element $x$. Note that for some variable $z_i$, it must be that $z_i^*=u$. We denote this as the an un-labelled arrow from $u$ to $z_i^*$.
The partial order on the pole-tuple defined earlier can be understood as follows. 
The point $z^*_i$ is said to be larger than $z^*_j$ if it can be reached from $z^*_j$ via a directed path.  The edges of the ordered graph, indicate the ``active'' pole-terms of the integrand. For example, consider the edge,
\begin{align}
    z_i^* \xrightarrow{x} z_j^* \qquad \Rightarrow \qquad \frac{1}{1-x\frac{z_i}{z_j}}.
\end{align}
If $z_j$ is integrated after $z_i$, then the above term only gives rise to the pole in $z_j$, not in $z_i$ because the pole in $z_i$ is outside the unit-circle. However, if $z_j$ is integrated before $z_i$ then it can be used to compute residue both in  $z_i$ and $z_j$. The contribution of the ordered pole-tuple must include all possible residues. It can be isolated as follows. Consider the contour $C_i$ which is the circle of $\epsilon$ radius around $z_i^*$. The contribution of the ordered-pole tuple is  the contour integral \eqref{general-int} where the contour for $z_i$ variable is $C_i$ rather than the unit circle.

Let us give an example of a pole-tuple in the simple setting of half-BPS index. Explicitly, the integral is given by
\begin{align}
    {\cal I}_N=\oint \frac{dz_i}{2\pi i (z_i-u)} \prod_{i\neq j}\frac{1-\frac{z_i}{z_j}}{1-a \frac{z_i}{z_j}}.
\end{align}
Consider the case $N=2$. We first pick the pole $z_1=u$. Note that because of the numerator $1-\frac{z_1}{z_2}$, there is no pole at $z_2=u$. The only pole for $z_2$ is at $z_1^* a=u a$. So this pole is labeled by the pole tuple $(u,ua)$. The associated directed graph is
\begin{align}
    u\to z_1^*\xrightarrow{a} z_2^*. 
\end{align}
We could instead first pick the $z_1$ pole at $a z_2$. Then the pole in $z_2$ is at $u$. This gives us the pole-tuple $(ua, u)$. 
The associated directed graph is
\begin{align}
    u\to z_2^*\xrightarrow{a} z_1^*. 
\end{align}
Both these pole contribute the same residue. As we will see, this is not a surprise. It is a very important and useful property of the ordered pole-tuples as we will explain shortly. 
To compute the integral we need to sum over the contribution of all ordered pole-tuples. This sum can be arranged as a sum over \emph{un-ordered} pole-tuples, where summand is the sum over all orderings of a given pole-tuple. 

Consider the pole-tuple that is obtained by another by permutation. Because of the permutation symmetry of the integral, the contribution of the differently ordered pole-tuple is equivalent to computing the $C_i$ integral but in the new order. However, due to Fubini's theorem, such a multi-variable integral does not depend on the order of integration and hence all orderings of a given pole-tuple contribute the same. This explains why the two ordered pole-tuples in the above examples gave the same residue.
As a result, the contour integration is expressed as sum over all \emph{unordered} pole-tuples, each taken with some preferred ordering, multiplied by $N!$. We construct this preferred order shortly but before that, let us verify the order independence of the contribution of ordered pole-tuples in some other examples. 

Consider the Hall-Littlewood index with $N=3$. Consider the ordered pole-tuple
\begin{align}
    u\to z_1^* \xrightarrow{a} z_3^* \xleftarrow{b} z_2^*.
\end{align}
This contributes zero, because as explained earlier, both of the two arrows are used to create pole for $z_3$ and not $z_1$ or $z_2$. As a result, the result of $C_i$ integration for variables $z_1$ and $z_2$ vanishes. Indeed, if we remove $z_3^*$ and the edges from the above directed graph, the graph disconnects. In particular, there is no pole-term that connects $z_1^*$ to $z_2^*$. Now consider the same tuple but with a different order
\begin{align}
    u\to z_1^* \xrightarrow{a} z_2^* \xleftarrow{b} z_3^*.
\end{align}
As remarked earlier, the arrow between $z_2^*$ and $z_3^*$ can used create poles for both $z_2$ and $z_3$. It turns out that both these residues are equal and opposite giving a vanishing total contribution. In this example, we see in an interesting way, that the ordering of the ordered pole-tuple does not matter.

\subsubsection{Preferred (partial) order}
We now construct the most convenient ordering on a given pole-tuple that will simplify the residue sum. We will choose it so that $N$ variable pole-tuple can be constructed from $N-1$ variable pole-tuple in an inductive way. In other words, we will pick the ordering $(z_1^*,  \ldots, z_{N-1}^*, z_N^*)$ on the pole-tuple such that omission of the last entry gives rise to the $N-1$ variable pole-tuple $(z_1^*,  \ldots, z_{N-1}^*)$. As a result, all the pole-tuples can be constructed inductively. Note that for this strategy to work, we don't really have to specify an order on for the tuple entries but rather we have to identify $z_N^*$. 
\begin{proposition}
    $z_N^*$ is a maximal element.
\end{proposition}
\noindent 
Let us explain. Consider the term in the denominator 
\begin{align}
\frac{1}{1-x\frac{z_i}{z_j}},    
\end{align}
for some $x\in P$. 
Let us assume that $j>i$ so that $z_j$ is integrated after $z_i$. In that case, as argued earlier, this term can only lead to the contribution of the pole in $z_j$ and not in $z_i$ because the pole in $z_j$ - caused by this term - is at $z_j^*= x^{-1} z_i$ and give that $z_i$ is un-integrated and on the unit circle, this pole is outside the unit circle of $z_j$. 
This discussion shows that if $z_N^*$ is a maximal element of the tuple, the active pole-terms involving $z_N$ are all used to construct the pole in $z_N$ and not in any other variable. The active pole-terms involving $z_N$ correspond to the arrows that end on $z_N^*$. If we remove $z_N^*$ and its arrows, the resulting directed graph must then be some ordered pole-tuple for ${\cal I}_{N-1}$ integral.  This shows us that,
\begin{proposition}\label{box-addition}
    Any contributing pole-tuple for ${\cal I}_N$ is constructed from that of ${\cal I}_{N-1}$ by adding an element such that it is a maximal element of the resulting $N$ pole-tuple.
\end{proposition}

This strategy effectively reduces the $N$ variable integral to a single variable integral. Denoting the integrand of ${\cal I}_N$ as $i_N$,
\begin{align}
    {\cal I}= \oint \prod_{i =1}^N \frac{dz_i}{2\pi i z_i} i_N &= \oint  \frac{dz_N}{2\pi i z_N} \Big( \oint \prod_{i =1}^{N-1} \frac{dz_i}{2\pi i z_i} i_{N-1}\Big) \notag\\
    &\times \prod_{i=1}^{N-1}(1-\frac{z_i}{z_N})(1-\frac{z_N}{z_i})  \frac{\prod_{y \in Z}(1-y \frac{z_i}{z_N})(1-y \frac{z_N}{z_i})}{\prod_{x\in P}(1-x \frac{z_i}{z_N})(1-x \frac{z_N}{z_i})}.
\end{align}
The bracket around $N-1$ variable indicates that, first the residue is taken in those variables according to the ordered-tuple $(z_1^*,\ldots, z_{N-1}^*)$. This residue is a function of $z_N$ due to the $z_N$ dependent terms on the second line. The maximal element $z_N^*$ of the $N$-tuple is identified as a pole of the $z_N$ dependent residue.

Now we are ready to describe the pole tuples for various index limits such as half-BPS, HL and BMN. 

\subsection*{Half-BPS index}
\begin{proposition}
    The Half BPS index is $N!$ times the contribution of the following  ordered pole-tuple.
    \begin{align}
        u\xrightarrow{}z_1^*\xrightarrow{a} z_2^* \xrightarrow{a} \ldots \xrightarrow{a} z_N^*.
    \end{align}
\end{proposition}
\noindent 
\begin{proof}
    This is proved straightforwardly with induction using the result of the previous subsection. The case of ${\cal I}_1$ is explicitly checked. 
    The inductive assumption that the above tuple is the only tuple contributing to ${\cal I}_N$. Using proposition \ref{box-addition}, we need to add an element to this ordered graph such that it is maximal and engineers a pole in the new variable. The only option is to add it to the right of $z_N^*$ with an arrow pointing to it from $z_N^*$. 
\end{proof}

\subsection*{Hall-Littlewood index}
\begin{figure}
    \centering
    \includegraphics[width=0.6\linewidth]{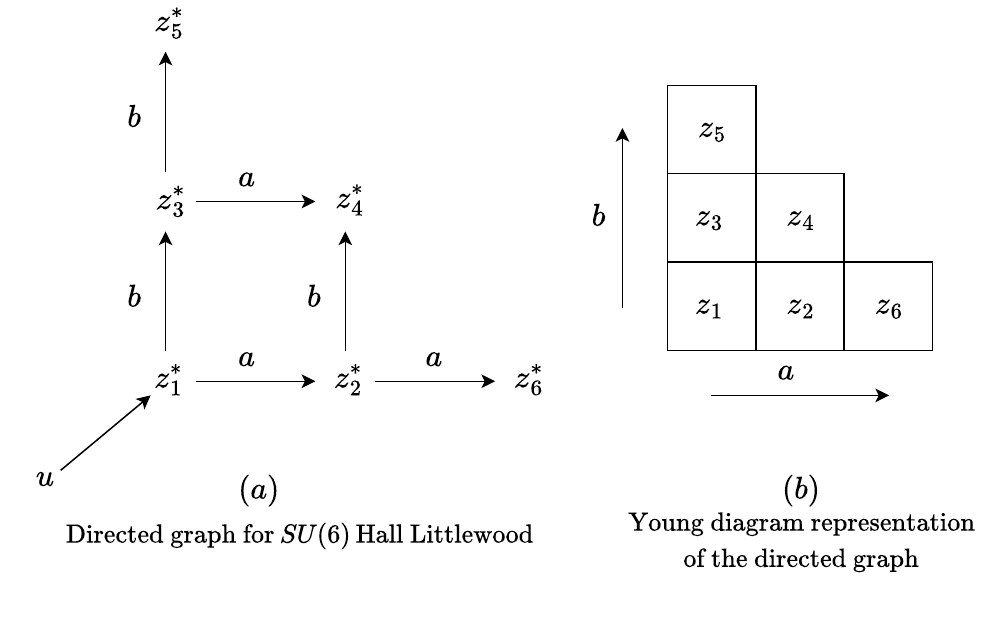}
    \caption{Young diagram representation of a SU(6) directed graph}
    \label{dr_gr_6}
\end{figure}

\begin{figure}
    \centering
    \includegraphics[width=0.7\linewidth]{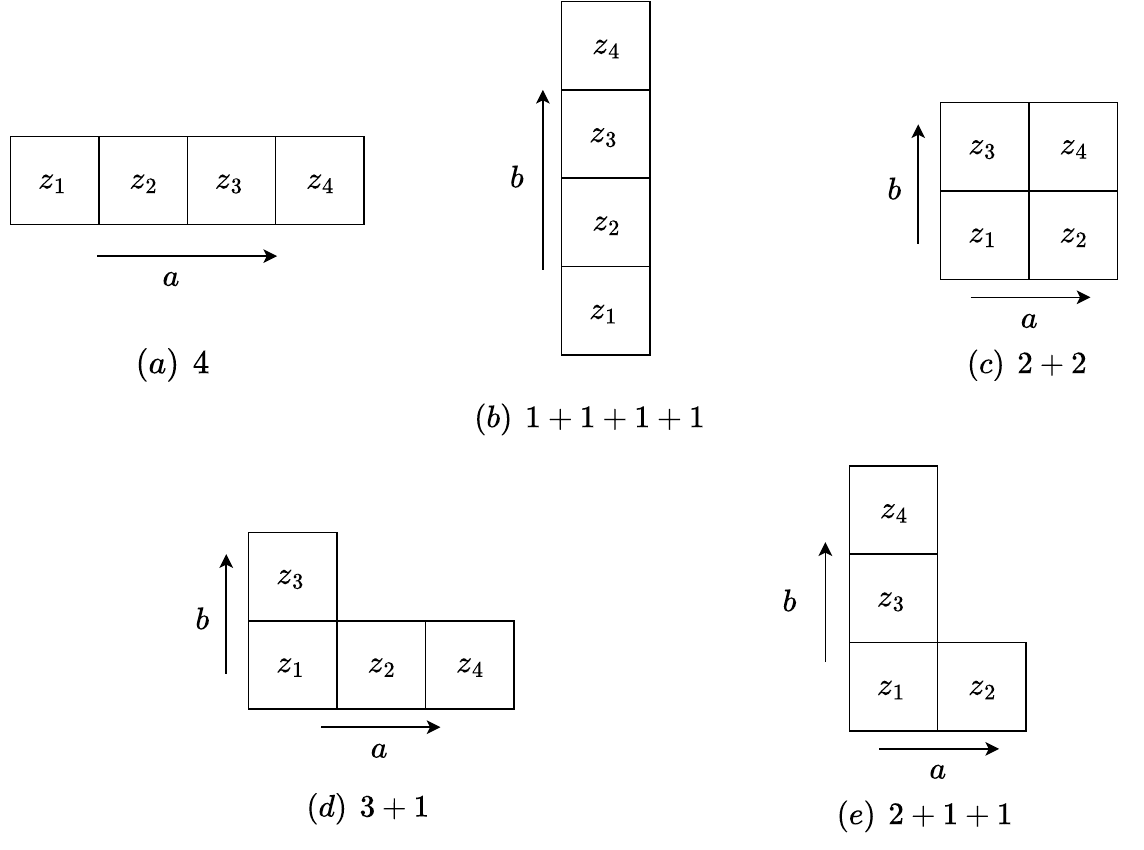}
    \caption{Young diagrams representing all possible poles contributing to the Hall-Littlewood index in SU(4) theory. }
    \label{yd4}
\end{figure}

\begin{proposition}
    The pole-tuples contributing to the Hall-Littlewood index ${\cal I}_N$ are shaped like Young diagram with $N$ boxes with $a$-arrows ($b$-arrows) pointing to the nearest right (upward) neighbor. The unlabeled arrow from $u$ is pointed to the first box. 
\end{proposition}
\noindent Figure \ref{dr_gr_6} gives an example of representation of directed graph as Young diagram. As an example, see all the directed graphs that contribute to the integral ${\cal I}_4$ in figure \ref{yd4}.
 
\begin{proof}
    The proof proceeds by induction. For $N=1$, the claim is explicitly checked. Given a Young diagram with $N$ boxes, we need to add another box such that it is maximal. The only choice is to add it in a way that results in a young diagram with $N+1$ boxes. All such additions correspond to non-zero residue for $z_{N+1}^*$ pole. 
\end{proof}
\noindent
Interestingly, the integral corresponding to the Hall-Littlewood index is almost the same as the one appearing in the computation of ``K-theoretic'' i.e. 5d instanton partition function in the $U(1)$ gauge theory. For $U(k)$ gauge theories, the integrand is similar as the Hall-Littlewood index except that it contains $k$ denominator factors i.e. $\prod_i (z-u_i)$. This allows one to use any of the $u_i$ to ``seed'' the pole-tuple. 
It was conjectured in \cite{Nekrasov:2003rj} that the poles are classified by a tuple of young diagrams, each seeded by one of the $u_i$'s. This was proved rigorously in \cite{Felder:2017rgg}. The proof presented there uses case-by-case analysis and hence is substantially lengthy. We obtain a short proof of the same by generalizing the inductive argument presented above to the $k$-tuple of Young-diagrams rather than a single one. 

\subsubsection{BMN index}
\begin{figure}
    \centering
    \includegraphics[width=0.7\linewidth]{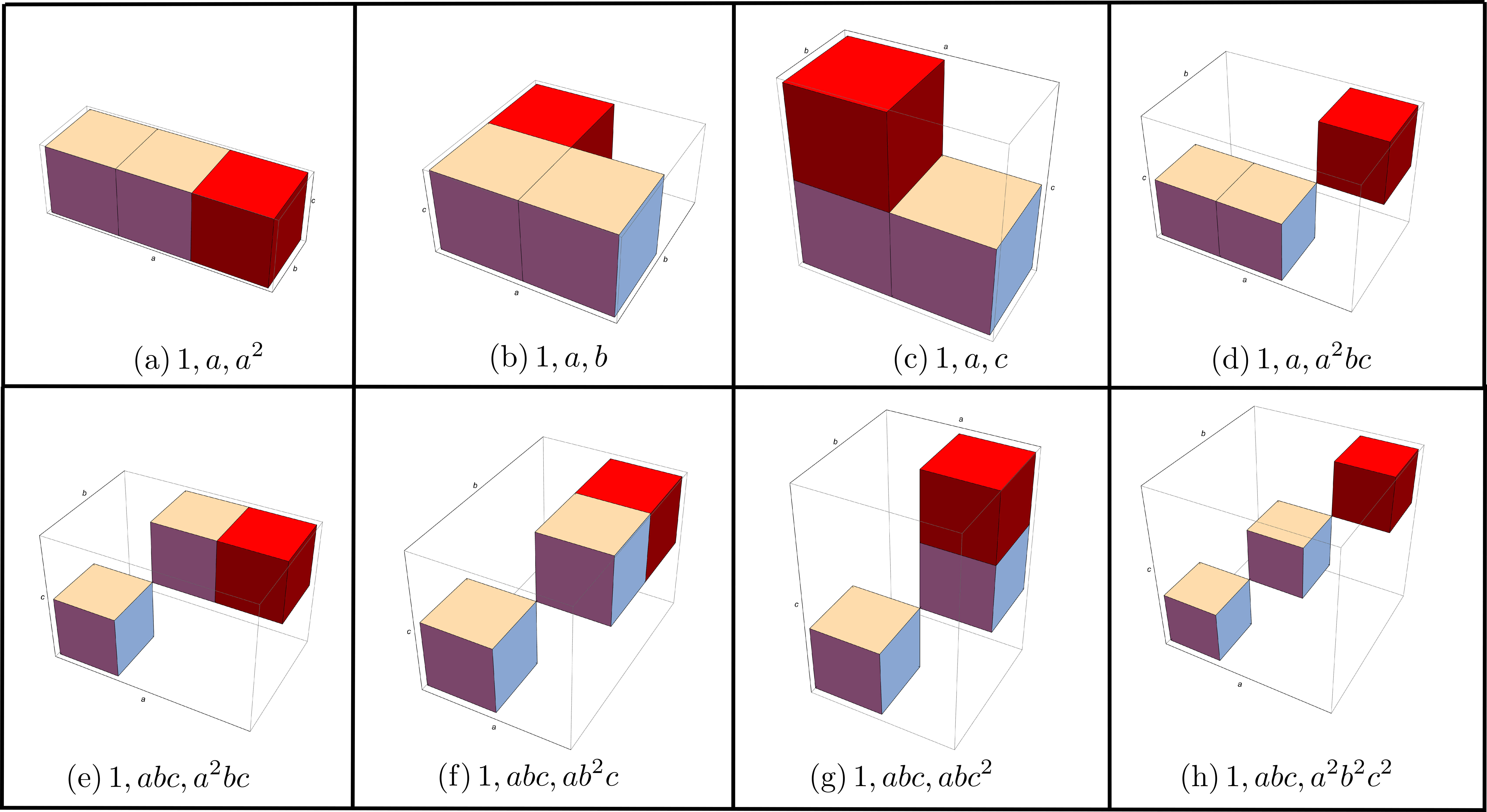}
    \caption{Poles of ${\cal I}_3$ in the BMN limit. }
    \label{su3}
\end{figure}

\begin{figure}
    \centering
    \includegraphics[width=0.3\linewidth]{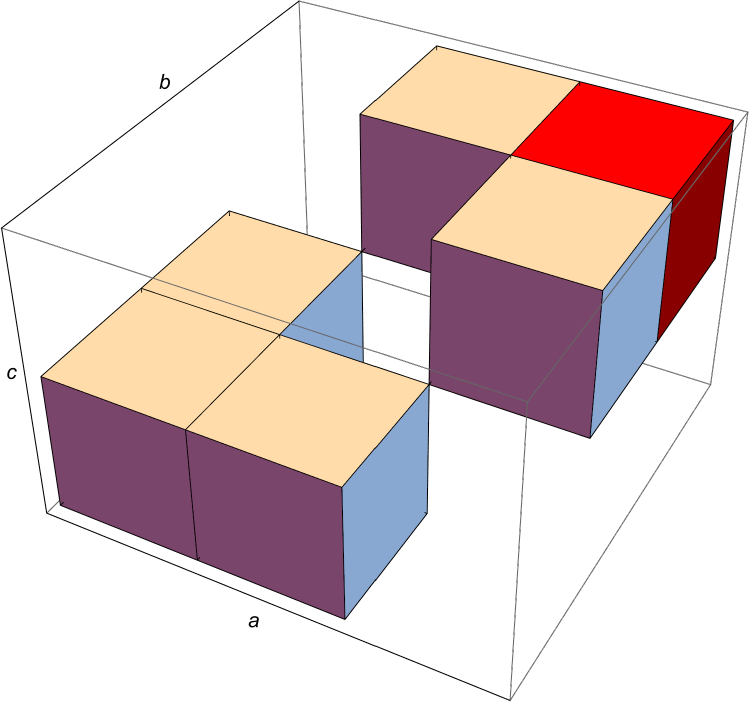}
    \caption{A  loop-like configuration of the pole for the BMN limit for $N=6$. It gives rise to a double pole in one of the integration variables. }
    \label{su6}
\end{figure}

The construction of pole-tuples that contribute to the integral is also done inductively. It is tempting to guess that it corresponds to three-dimensional Young diagrams but it is not so. This is easy to see already for $N=2$. The only pole-tuple for $N=1$ is $z_1^*=u$. For $N=2$, we need to add $z_2^*$ such that it is not less than $z_1^*$. Of course, it also needs to correspond a pole in the $z_2$ integral, meaning that there must be an arrow between $z_1^*$ and $z_2^*$, so in fact $z_2^*$ must be greater than $z_1^*$. There are four choices corresponding to the four elements of $P$. 
\begin{align}
    z_2^* \in \{ua, \quad ub, \quad uc, \quad uabc\}
\end{align}
The first three of these choices correspond to the three three-dimensional Young diagram with two boxes, however, the fourth choice corresponds to adding the new box such that it is touching the previous box at a vertex. This does not correspond to a three-dimensional Young diagram. Nevertheless, we proceed with induction to construct pole-tuples for any $N$ that we are interested in. As an example, we have listed the contributing  configuration of boxes in figure \ref{su3} for  $N=3$. A novel, loop-like configuration is encountered for $N=6$ as shown in figure \ref{su6}.

In the next section, we present the result of our computation of the BMN index for $N=2, \ldots, 6$. To avoid clutter, we have set $a=b=c=t^2$ in the final answer. 
We have also checked these results by computing the integral by first expanding the integrand in $t$ and integrating each term of the expansion separately.  

\subsection{BMN index of SU(2) to SU(6) SYM}
In this section we give the BMN index of SU($N$) SYM for $N=2,\ldots, 6$ in the closed form computed using the residue integral. Recall that $t$ is a fugacity for the charge $j\equiv 2(Q_1+Q_2+Q_3)+3(J_1+J_2)$

\begin{itemize}
    \item SU(2) BMN index
    \begin{align}\label{su2full}
    &I_{SU(2)}=(1-t^2)^3\nonumber\\
    &\frac{1 \!+\! 3 t^2 \!+\! 12 t^4 \!+\! 20 t^6 \!+\! 42 t^8 \!+\! 48 t^{10} \!+\! 75 t^{12} \!+\! 66 t^{14} + 
 81 t^{16} \!+\! 55 t^{18} \!+\! 54 t^{20} \!+\! 27 t^{22} \!+\! 19 t^{24} \!+\! 6 t^{26} \!+\! 3 t^{28}}{(1 - t^8)^3 (1 - t^{12})}
\end{align}
    \item SU(3) BMN index
\begin{align}
    I_{SU(3)} = & \ (1\!+\!5 t^2\!+\!24 t^4\!+\!82 t^6\!+\!245 t^8\!+\!628 t^{10}\!+\!1444 t^{12}\!+\!3013 t^{14}\!+\!5806t^{16}\!+\!10431 t^{18} \nonumber \\
&\!+\!17628 t^{20}\!+\!28174 t^{22}\!+\!42820 t^{24}\!+\!62104 t^{26}\!+\!86268 t^{28}\!+\!115035 t^{30}\!+\!147595 t^{32}\nonumber \\
&\!+\!182506 t^{34}\!+\!217831 t^{36}\!+\!251223 t^{38}\!+\!280233 t^{40}\!+\!302516 t^{42}\!+\!316224t^{44}\!+\!320131 t^{46}\nonumber \\
&\!+\!313920 t^{48}\!+\!298116 t^{50}\!+\!274103 t^{52}\!+\!243875t^{54}\!+\!209819 t^{56}\!+\!174384 t^{58}\!+\!139852 t^{60} \nonumber \\
&\!+\!108049 t^{62}\!+\!80282t^{64}\!+\!57228 t^{66}\!+\!39035 t^{68}\!+\!25393 t^{70}\!+\!15687 t^{72}\!+\!9153 t^{74}\!+\!5013 t^{76} \nonumber \\
&\!+\!2553 t^{78}\!+\!1197 t^{80}\!+\!507 t^{82}\!+\!189 t^{84}\!+\!60 t^{86}\!+\!15 t^{88}\!+\!3 t^{90} ) \nonumber \\
&\times \frac{(1-t^2)^5 (1-t^4)^3 }{(1-t^8)^3 (1-t^{10})^2 (1-t^{12}) (1-t^{14})^3 (1-t^{18})}~,
\end{align}
\item SU(4) BMN index
\begin{align}
I_{SU(4)}= & \ (1+8 t^2+50 t^4+234 t^6+\cdots+71 t^{222}+11 t^{224}+t^{226}) \nonumber \\
&\times \frac{(1\!-\!t^2)^8 (1\!-\!t^4)^8}{(1\!-\!t^8)^3 (1\!-\!t^{10})^2 (1\!-\!t^{12}) (1\!-\!t^{14})^3 (1\!-\!t^{16})^5 (1\!-\!t^{18}) (1\!-\!t^{20})^3 (1\!-\!t^{24})}~.
\end{align}

\item SU(5) BMN index
\begin{align}
&I_{SU(5)}=\left(1+13t^2+108t^4+\cdots  + 146 t^{450} + 11 t^{452}\right) \nonumber \\
&\times \frac{\left(1-t^{2}\right)^{13} \left(1-t^4\right)^{11}\left(1-t^6\right)^{3}}{
   \left(1-t^{12}\right) \left(1-t^{14}\right)^3
   \left(1-t^{16}\right)^5
   \left(1-t^{18}\right)^4 \left(1-t^{20}\right)^3 \left(1-t^{22}\right)^5 \left(1-t^{24}\right)
   \left(1-t^{26}\right)^3 \left(1-t^{30}\right)}.
\end{align}
\item SU(6) BMN index
\begin{align}
&I_{SU(6)}=\frac{(1 + 14 t^2 + 131 t^4 + 933 t^6+\cdots+846 t^{818} + 51 t^{820} - 14 t^{822} - 5 t^{824} - t^{826})}{\left(1-t^8\right)^4 \left(1-t^{10}\right)
   \left(1\!-\!t^{14}\right)^4}\nonumber\\
&\frac{\left(1\!-\!t^2\right)^{14} \left(1\!-\!t^4\right)^{20} \left(1\!-\!t^6\right)^7
   \left(1\!-\!t^{12}\right)^3}{\left(1\!-\!t^{16}\right)^5 \left(1\!-\!t^{18}\right)^6
   \left(1\!-\!t^{20}\right)^4 \left(1\!-\!t^{22}\right)^5 \left(1\!-\!t^{24}\right)^7
   \left(1\!-\!t^{26}\right)^3 \left(1\!-\!t^{28}\right)^5 \left(1\!-\!t^{30}\right)
   \left(1\!-\!t^{32}\right)^3 \left(1\!-\!t^{36}\right)}
\end{align}
\end{itemize}
\begin{figure}
    \centering
    \includegraphics[width=1\linewidth]{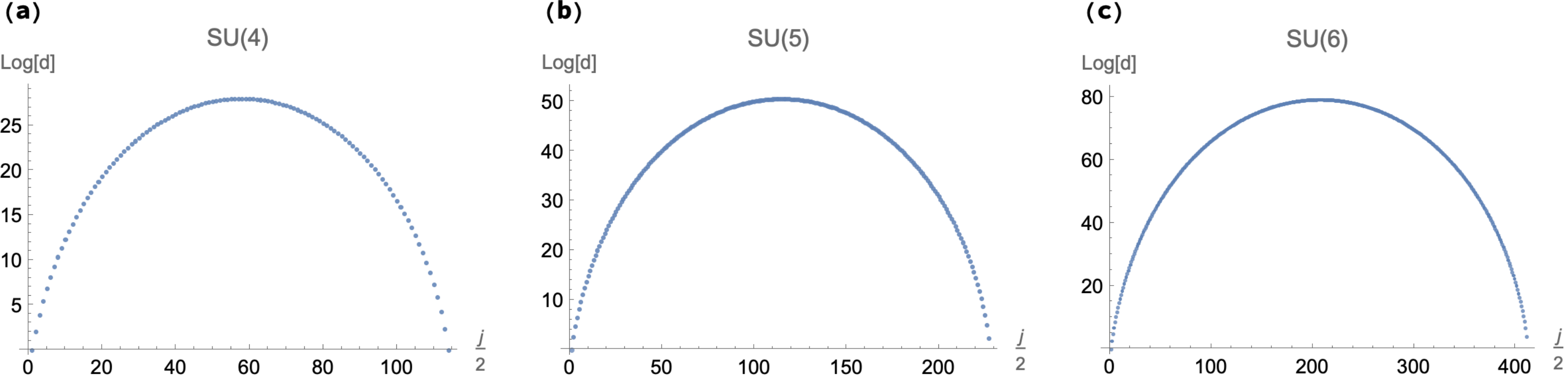}
    \caption{Logarithm of the absolute values of the coefficients in the numerators of the BMN indices for SU(4), SU(5), and SU(6).}
    \label{fig:bmn index coeff}
\end{figure}
Note the existence of the denominator factors $(1-t^{6N})(1-t^{6N-4})^3$ is all of the above examples. We will discuss these factors further in the next section. 
Even though we have explicitly computed them, the numerators for the SU(4), SU(5), and SU(6) indices are omitted in the equations above to avoid clutter. Instead,  in Figure \ref{fig:bmn index coeff}, we plot the logarithm of the absolute values of the coefficients in the parentheses of the numerators of the BMN indices, \( \log |d_j| \), for SU(4), SU(5), and SU(6) theories, where the coefficients are given by
\begin{align}
	\sum_j d_j t^{j}.
\end{align}  
The peak occurs at the midpoint of the series.  

Qualitatively, the curves in Figure \ref{fig:bmn index coeff} all exhibit a similar overall shape. Moreover, with the exception of a few terms in the SU(6) BMN index numerator, every coefficient is positive. It would be valuable to develop a quantitative characterization of these coefficient distributions at large
$N$ to get precise information about the entropy growth in the BMN sector.

\subsection{General structure of the index}\label{struc}
A partition function over a ring takes the form
\begin{align}\label{ring-deco}
    Z=\frac{P}{\prod_i (1-b_i)},
\end{align}
where $b_i$'s are monomials in the fugacities and $P$ is a polynomial in the fugacities with positive integer coefficients. This decomposition gives an intuitive way of understanding the structure of the ring. The monomials $b_i$ are the contributions of a certain ``primitive'' subset of ring generators and the polynomial numerator $P$ encodes the relations between them. The case $P=1$ corresponds to a ring that is freely generated by associated generators.

The superconformal index is a partition function over a CFT Hilbert space with $(-1)^F$ insertion. In CFT, the states in the Hilbert space are in one-to-one correspondence to operators. Operators can be multiplied giving the Hilbert space a ring structure. This makes  the index a supersymmetric partition function over a ring. As a result, it admits the decomposition \eqref{ring-deco} with two differences. In the case of the index, the denominator factors could come from only bosonic generators and the polynomial $P$ can have negative integer coefficients, due to $(-1)^F$ insertion. 

Now let us get back to the discussion of the full-BMN index computed in section \ref{pole res}. In this case, the residue at the pole-tuple $(1,abc, \ldots, (abc)^N)$ is
\begin{align}\label{fff}
	&(-1)^{N-1}\frac{1-f}{1-f^N} \prod_{k=1}^{N-1} \frac{(1 - f^k ab)(1 - f^k bc)(1 - f^k ca)}{(1 - f^k a)(1 - f^k b)(1 - f^k c)}, \qquad f\equiv abc.
\end{align}
The full answer is obtained by summing over all the residues. However, the factor $(1-f^N)(1-f^{N-1}a)(1-f^{N-1}b)(1-f^{N-1}c)$ are still present in the denominator of the sum. After setting $a=b=c=t^2$, they correspond to bosonic generators that contribute $t^{6N}, t^{6N-4}, t^{6N-4}, t^{6N-4}$ respectively. The presence of these denominator terms can be confirmed from the closed form expression of the BMN index for $N=2,\ldots, 6$ presented in section \ref{pole res}.

The graviton index is difficult to evaluate in closed form in general, however, the analysis of the denominator factors can  be performed. 
The graviton generators contributing to the BMN index for SU($N$) theory are listed in equation \eqref{bmnsusn}. They arise from \( S_k \) multiplets for \( k = 2, \dots, N \). The bosonic generators are $|k\rangle$ and $\bar{\mathcal{Q}}_+^m \bar{\mathcal{Q}}_+^n |k\rangle$ with multiplicity $(k+1)(k+2)/2$ and $k(k+1)/2$ respectively. They contribute,  
\begin{align}
	|k\rangle \to t^{2k}, \quad \bar{\mathcal{Q}}_+^m \bar{\mathcal{Q}}_+^n |k\rangle \to t^{2k+4}.
\end{align}
Here we have set \( a = b = c = t^2 \) for simplicity. If all of them are primitive generators then the denominator factor would be   
\begin{align}\label{grav tower}
	\prod_{k=2}^{N} (1 - t^{2k})^{\frac{(k+1)(k+2)}{2}} (1 - t^{2k+4})^{\frac{k(k+1)}{2}}.
\end{align}
Even if only a subset of generators are primitive, it is clear from this discussion that the monomials $b_i\in\{t^{4}, t^6,\ldots, t^{2N+4}\}$. In particular, the maximum power of $t$ that appears in the denominator factor is $t^{2N+4}$.

From the discussion of the graviton index, it is clear that the monomials $f^N$, $f^{N-1}a$, $f^{N-1}b$ and  $f^{N-1}c$ can not result from the bosonic generator that is a graviton operator. In fact the bosonic generators can be explicitly identified from their charges and they are ${\rm Tr}[f_{++}^N],{\rm Tr}[f_{++}^{N-1}\bar \phi_1],{\rm Tr}[f_{++}^{N-1}\bar \phi_2],{\rm Tr}[f_{++}^{N-1}\bar \phi_3]$ respectively.
Notably, the field strength \( f_{++} \) plays an important role in distinguishing towers in the full BMN  index from those in the graviton sector. 
In \( \mathcal{N}=4 \) SYM, the only bosonic letter apart from the chiral scalars \( \bar{\phi}^i \) is \( f_{++} \). Its presence appears to be a sufficient for hosting non-graviton tower structures. Without \( f_{++} \), single-trace bosonic operators can always be matched to graviton operators with the same charge, formed by chiral scalars or scalars dressed with derivatives.  
Interestingly, all known 1/4-BPS and 1/8-BPS sectors lack \( f_{++} \) and are expected not to contain non-graviton cohomologies due to ${\cal O}(1)$ scaling of the entropy in the large $N$ limit \cite{Kinney:2005ej, Chang:2023ywj}.\footnote{One might ask whether the presence of \( f_{++} \) is a necessary condition for a consistently truncated sector to host non-graviton operators. The answer is no. See Section \ref{1ltrun} for other 1-loop truncations that do not contain $f_{++}$, but has non-graviton states in it.}

Let us look more closely at the BMN index and the BMN graviton index for SU(2) SYM.

\subsection{SU(2) example}\label{example}\label{gravtower}
We explicitly compute the SU(2) BMN graviton index, following the procedure outlined in Section \ref{gravind} (for details, see Appendix C of \cite{Choi:2022caq}). The generators correspond to the single-graviton operators in the \(S_2\) multiplet. Since all fields can be treated as diagonal matrices, we only consider Cartan generators, where the letters \(\{\bar{\phi}^i, \psi_{i+}, f_{++}\} \in \Phi\) take the form:  
\begin{align}
	\Phi = \begin{pmatrix} \varphi &0\\0 &-\varphi\end{pmatrix},
\end{align}
where \(\varphi\) represents eigenvalues \(x, y, z, \psi_1, \psi_2, \psi_3, f\) for each letter. Here, \(\psi_{1,2,3}\) are Grassmann variables, while \(x, y, z, f\) are bosonic.

In terms of eigenvalues, the graviton operators in the \(S_2\) multiplet are expressed as:  
\begin{align}\label{BMNsggrav}
	x^2~,~y^2~,~z^2~,~xy~,~yz~,~zx~,~xf+\psi_2\psi_3~,~yf+\psi_3\psi_1~,~zf+\psi_1\psi_2~, \nonumber \\
	y \psi_1~,~ z\psi_1~,~ z\psi_2~,~ x\psi_2~,~ x\psi_3~, ~y\psi_3~,~ x\psi_1-y\psi_2~,~ y\psi_2-z\psi_3~.
\end{align}
From these generators, the graviton index is computed as:  
\begin{align}
	I_{\text{grav}} = \frac{\text{numerator}}{(1-a^2)(1-b^2)(1-c^2)(1-a^2bc)(1-ab^2c)(1-abc^2)},
\end{align}
where the numerator is a polynomial of \(a, b,\) and \(c\). Setting \(a = b = c = t^2\), it simplifies to:
\begin{align}
	\text{numerator} = 1\!&+\!3t^4\!-\!8t^6\!-\!6t^{10}\!+\!10t^{12}\!+
	\!9t^{14}\!-\!9t^{16}\!\nonumber\\
	&+\!16t^{18}\!
	-\!18t^{20}\!-\!3t^{22}\!+\!t^{24}\!-\!3t^{26}\!+\!9t^{28}\!-\!2t^{30}\!+\!3t^{32}\!-\!3t^{34}.
\end{align}
Notably, terms like \(1-ab\), \(1-bc\), and \(1-ca\) do not appear in the denominator, as they are canceled in the numerator due to polynomial relations such as \((xy)^2 = x^2 y^2\) and its cyclic permutations.

Subtracting the BMN graviton index from  the full BMN index \eqref{su2full} yields the BMN index over non-graviton states. Curiously, it has a very simple expression. 
\begin{align}
    I_{SU(2)}-I_{SU(2)}^{grav}=-t^{24}\frac{(1-t^2)^3}{(1-t^8)^3(1-t^{12})}
\end{align}
recall that $t$ is a fugacity for the charge $j=2(Q_1+Q_2+Q_3)+3(J_1+J_2)$.
This expression gives remarkable insights into the structure of non-graviton states of SU(2) SYM. First of all, the leading contribution $-t^{24}$ is the contribution of the lightest non-graviton \emph{a.k.a.} fortuitous state $O_0$. We will refer to this state as the core black hole. 
See \cite{Choi:2023znd} to see the explicit expression of this state in terms of BMN letters. The other factors have the following interpretation. 
\begin{itemize}
    \item $(1-t^2)^3$ factor: As described earlier, the BMN index is closed under PSU$(1|3)$ subalgebra. This factor is precisely the contribution of superconformal descendants with respect to this alebra i.e. decsendants under the action of  $\bar{\mathcal{Q}}_{+}^{m}$.
    \item $(1-t^8)^3$ factor: Note that $t^8= t^{2N+4}$ for $N=2$ i.e. its contribution matches that of the heaviest graviton operator. The appearance of this factor in the denominator of the non-graviton index suggests that multiplying the core black hole state $O_0$ with any power of the heaviest graviton operator results in the operator that is BPS at 1-loop. This is reminiscent of the grey galaxy solution recently constructed in \cite{Kim:2023sig,Bajaj:2024utv}. It corresponds to a gas of gravitons remotely circling a black hole.
    \item $(1-t^{12})$ factor: We feel that this is the most curious factor. One would be tempted to interpret it also as resulting from graviton excitations away from the black-hole and hence being part of the ``grey-galaxy'' solution. However, the power of $t$ is not compatible with any of the graviton generators. It corresponds to a tower of operators
    \begin{align}\label{towersu2}
        O_n= \text{Tr}[f_{++}^2]^nO_0+n\text{Tr}[f^2]^{n-1}\text{Tr}[f_{++} \xi]+\frac{2n^2+n}{3}\text{Tr}[f_{++}^2]^{n-1}\chi,~~~~n\geq 1
    \end{align}
    where $O_0$ is the non-graviton cohomology at $j=24$. $\xi$ and $\chi$ are given in terms of free BPS letters (which are not important in this paper. For the expression of these terms see \cite{Choi:2022caq}, for example.) 
    As one can see from the first term in the RHS of \eqref{towersu2}, the tower is given by the Tr$f_{++}^2$ operator as argued in the previous section. 
\end{itemize}

It would be instructive to carry out a similar analysis for higher values of $N$. However, computation of the graviton index is very difficult and requires large amount of computational resources. We hope to overcome these hurdles and push the calculation of non-graviton index to higher values of $N$ in the future. 
Nevertheless, as explained in the previous section, the non-graviton denominator structure i.e. of the type $(1-t^{12})$ in the SU(2) example, can still be inferred from the denominator of the full index.  
From SU(2) to SU(6), we observe the existence of a non-graviton bosonic tower spanning from \( j = 2N+6 \) to \( 6N \), with one exception: \( j = 6N-2 \).  
This exception arises because there is no single-trace bosonic operator with charge \( j = 6N-2 \).  The residue at the pole-tuple $(1, abc, \ldots, (abc)^N)$ suggests that this is true for all values of $N$.

On the dual side, existence of this tower suggests an interesting Fock space of bosonic excitations that are not gravitons. It is tantalizing to speculate that it corresponds to the bosonic excitations of the black-hole itself or to brane configurations that explore the near horizon limit of the black hole. We do not have any concrete suggestions on this matter and hope to further explore this issue in the future.

\section{1-loop truncation and the S duality of $\mathcal{N}=4$ SYM}\label{sospbmn}
In this section, we will discuss some consequences of the conjecture. The conjecture states that the exact cohomology ring is isomorphic to the 1-loop cohomology ring with the isomorphism that preserves 
charges. We can identify several sub-rings at the level of  1-loop cohomology using truncation based on letters. We call such sub-rings, letter-based truncations ${\cal T}$. According to the conjecture, the exact cohomology must also have subrings that correspond to these truncations. In particular, ${\cal N}=4$ SYM with gauge group $G$ must have these subrings at both, weak and strong coupling. However, according to S-duality, this theory at strong coupling is described by  ${\cal N}=4$ SYM with gauge group $\hat G$ at weak coupling where $\hat G$ is Langlands dual of $G$. Question: is  the subring ${\cal T}$ for $G$ gauge group isomorphic to  the subring ${\cal T}$ for $\hat G$ gauge group. If this is not the case and \emph{if the conjecture is true} then this implies that they must exist new \emph{non-letter-based} subrings $\tilde {\cal T}$ of the 1-loop cohomology  of the $\hat G$ theory that  matches with the ${\cal T}$ truncation of the $G$ theory and vice versa. Non-existence of such novel $\tilde {\cal T}$ subrings in 1-loop cohomology can then potentially falsify the conjecture. 

We will indeed find that it is not the case that the truncation ${\cal T}$ for $G$ gauge group isomorphic to  the truncation ${\cal T}$ for $\hat G$ gauge group by computing the index on both sides. If we take $G=\text{SU($N$)}$ then $\hat G= \text{SU($N$)}/{\mathbb Z}_n$. Because the gauge algebra is the same on both sides, the index computation over any truncation ${\cal T}$ is the same for both the gauge groups. In order to see the mismatch we have to pick the S-dual pairs $G=\text{Sp($N$)}$ and $\hat G=\text{SO($2N+1$)}$.  This S-duality is non-trivial only for $N\geq 3$. This is because Sp(1)$\sim$ SO(3) and Sp(2)$\sim$ SO(5). In the rest of the subsection we consider the case $N=3$ i.e. the S-duality between Sp(3) and SO(7) ${\cal N}=4$ SYMs. We will take ${\cal T}$ to be the BMN truncation although a similar analysis can be done for any other letter-based truncation.

Before we turn to calculation of the BMN index for the S-dual pairs, let us discuss the graviton spectrum across S-duality. Interestingly, the graviton spectrum is isomorphic in S-dual theories.\footnote{In general, identifying the dual of an arbitrary operator in the dual theory is a challenging task.}  This is because, the graviton operators can be effectively treated as elements of the chiral ring, where  fields are are valued in the Cartan subalgebra and mutually commute.  
Since the Cartan subalgebras of Sp($N$) and SO($2N+1$) are identical, they lead to the same graviton spectrum. Below, we provide a brief explanation of this correspondence.\footnote{Based on this argument, one might wonder whether the graviton Hilbert space of SO(2N) is also isomorphic to that of Sp($N$) or SO($2N+1$). However, this is not the case due to the presence of Pfaffian operators in SO(2N), which contribute additional elements to the graviton spectrum that cannot be generated from the chiral primary operators alone \cite{Witten:1998xy}. To correctly account for graviton states in SO(2N), one must include Pfaffian operators in addition to the \( S_n \) multiplets.}

In the case of Sp($N$), the Cartan subalgebra consists of $N$ generators represented as $2N \times 2N$ diagonal matrices. For example, the $k$-th generator can be expressed as:
\begin{align}
    H_k = \text{diag}\big(\underbrace{0, \dots, 0}_{2k-2}, i, -i, \underbrace{0, \dots, 0}_{2N+1-2k} \big).
\end{align}
For SO($2N+1$), the Cartan subalgebra also contains $N$ generators, but these are block-diagonal antisymmetric matrices, with \( 2 \times 2 \) blocks corresponding to rotations in distinct planes. The additional row and column associated with the \( (2N+1) \)-th dimension are zeros. For a general $k$, the $k$-th generator can be written as:
\begin{align}
    H_k = \text{diag}\big(\underbrace{0, \dots, 0}_{2k-2}, \begin{pmatrix}
0 & 1 \\
-1 & 0
\end{pmatrix}, \underbrace{0, \dots, 0}_{2N+1-2k} \big).
\end{align}
Since $N$ Cartan generators in Sp and SO are related by similarity transformations, the subalgebras are identical. Therefore, the graviton spectrum for these two theories necessarily coincide. 
The single-graviton spectrum for Sp and SO can be organized into $S_n$ multiplets, as in the $SU$ case.  
Note that, however, $S_{n}$ multiplets with $n=2k+1$ vanish for both SO and Sp gauge groups due to the antisymmetric properties of their generators. 
For instance, this can be shown for a Sp/SO generator $X$ as follows:  
\begin{align}
    \text{Tr}[X^{2k+1}] = \text{Tr}[{X^T}^{2k+1}] = -\text{Tr}[X^{2k+1}],
\end{align}  
where the first equality uses the invariance of the trace under transposition, and the second equality relies on the properties of the generators: for SO($2N+1$), \( X^T = -X \), and for Sp($N$), \( \Omega X = -X^T \Omega \), where $\Omega$ is the symplectic form given by $\Omega = \begin{pmatrix} 0 & I \\ -I & 0 \end{pmatrix}$.

\subsection{SO($2N+1$) and Sp($N$) BMN index}\label{sdnongr}
A highly nontrivial test of $S$-duality between SO/Sp theories is the matching of the two superconformal indices, which has been verified for various gauge group ranks and sufficiently large charges \cite{Gadde:2009kb, Spiridonov:2010qv}.

In this section we compute the BMN index for these theories in order to see if they match. We will set up the calculation for general $N$, estimate it in the large $N$ limit.  In the next subsection we will focus on the case $N=3$. 
The character of the adjoint representation for Sp($N$) and SO($2N+1$) gauge group respectively is 
\begin{align}
    \chi_{Sp(N)}^{\rm adj}&= \sum_{i< j} \Big(\frac{z_i}{z_j}+\frac{z_j}{z_i}+z_i z_j+\frac{1}{z_i z_j}\Big)+\sum_i z_i^2\notag\\
    \chi_{Sp(N)}^{\rm adj}&= \sum_{i< j} \Big(\frac{z_i}{z_j}+\frac{z_j}{z_i}+z_i z_j+\frac{1}{z_i z_j}\Big)+\sum_i z_i.
\end{align}
Accordingly, the BMN index is given by the  contour integrals,
\begin{align}\label{bmnsospind2}
    I_{SO(2N+1)} &= \frac{(\kappa/2)^{N}}{N!} \oint \prod_{i=1}^N \frac{dz_i}{2\pi i z_i} \prod_{j< i}^N F\left(\frac{z_i}{z_j}\right) F\left(\frac{z_j}{z_i}\right) F(z_{i}z_j) F\left(\frac{1}{z_iz_j}\right) \prod_{i=1}^N F(z_{i}) F\left(\frac{1}{z_i}\right), \nonumber\\
    I_{Sp(N)} &= \frac{(\kappa/2)^{N}}{N!} \oint \prod_{i=1}^N \frac{dz_i}{2\pi i z_i} \prod_{j< i}^N F\left(\frac{z_i}{z_j}\right) F\left(\frac{z_j}{z_i}\right) F(z_{i}z_j) F\left(\frac{1}{z_iz_j}\right) \prod_{i=1}^N F(z_{i}^2) F\left(\frac{1}{z_i^2}\right),
\end{align}
where the function \( F(z) \) is given by:
\begin{align}
    F(z) = \frac{(1 - z)(1 - abz)(1 - bcz)(1 - caz)}{(1 - abcz)(1 - az)(1 - bz)(1 - cz)}.
\end{align}
If we replace $a=b=c=t^2$ then the term in the denominator corresponding to bosonic generator with maximum $t$ weight comes from the residue at the pole-tuple
\begin{align}
    z_1 = abc z_2, \quad z_2 = abc z_3, \quad \dots, \quad z_{N-1} = abc z_N,
\end{align}
just as in the case of SU($N$) theory. It leads to the highest-degree term in the denominator $1 - (abc)^{2N} = 1 - t^{12N}.$
This corresponds to the non-graviton tower with maximal power generated by the operator \( \text{Tr} f_{++}^{2N} \), which arises from the highest-degree power of the field strength \( f_{++} \) which is single-trace. 

Following the argument given in Sections \ref{struc} and \ref{gravtower}, the presence of the \( (1 - (abc)^{2N}) \) tower provides evidence for the existence of non-graviton states in SO/Sp theories.  Let us review the argument. 
The single-graviton operators in SO($2N+1$) and Sp($N$) consist of the multiplets \( S_{2k} \) with \( k \leq N \).   
The bosonic graviton states within the \( S_{2k} \) multiplets contribute to the index with fugacities \( t^{4k} \) and \( t^{4k+4} \). 
This implies that the highest tower constructed purely from gravitons is generated by the terms \( t^{4N+4} \).

One can also perform a saddle point analysis for the SO/Sp BMN index given in \eqref{bmnsospind2}.  
In fact, the analysis is similar to that of SU($N$), and the resulting saddle point distribution closely resembles the SU($N$) case:  
\begin{align}\label{sosp saddle}  
    \rho(\alpha) = \frac{3}{2\pi^3}(\pi^2 - \alpha^2), \quad 0 \leq \alpha \leq \pi.  
\end{align}
This is because the integrand of \eqref{bmnsospind2} has similar structure to that of \eqref{int0}. Specifically, in the large \( N \) limit, the two-body potential in the exponent can be related to the two-body potential of the SU($N$) integral as follows:\footnote{There are also one-body potential terms in \eqref{bmnsospind2}. However, these terms are subleading in the large \( N \) limit, as the number of two-body interactions scales as \( O(N^2) \), which is significantly larger than the \( O(N) \) contribution from the one-body terms.}  
\begin{align}
    V(\alpha_i - \alpha) \approx V_{SU}(\alpha_i - \alpha) + V_{SU}(\alpha_i + \alpha).
\end{align}
As a result, the SO/Sp saddle \eqref{sosp saddle} satisfies the saddle point equation because it can be rewritten in terms of the SU($N$) saddle point equation with \( \rho_{SU}(\alpha) \) given by \eqref{su saddle}:
\begin{align}\label{saddle sosp}
    \int_{0}^{\pi} d\alpha\, \rho(\alpha) V'(\alpha_i - \alpha) = \int_{-\pi}^{\pi} d\alpha\, \rho_{SU}(\alpha) V_{SU}'(\alpha_i - \alpha) = 0.
\end{align}
The free energy in the large \( N \) limit is given by:
\begin{align}
    \log I = -\frac{1}{2} \frac{3(2N)^2}{2\pi^2} \Delta_1 \Delta_2 \Delta_3,
\end{align}
where $\Delta_I$ are defined by $a=e^{-\Delta_1},~~b=e^{-\Delta_2}~~c=e^{-\Delta_3}.$
The free energy is precisely twice that of SU($N$). Consequently, the entropy is reduced by half. This reduction is natural since, in the large \( N \) and large 't Hooft coupling limit, \( \mathcal{N}=4 \) SYM theories with gauge groups SO($2N+1$) and Sp($N$) are dual to type IIB supergravity on the orbifold geometry AdS\(_5 \times S^5/\mathbb{Z}_2\). The volume of the internal space is reduced by a factor of two, leading to a corresponding reduction in entropy. (A similar analysis for the full index was carried out in \cite{Amariti:2020jyx,Choi:2023tiq}.)

At strict large $N$ limit, the index for SO/Sp theory is equivalent to the graviton index. Therefore, it can be obtained by summing over all the contributions from $S_{2k}$ multiplets where $k\in \mathbb{N}$, given by
\begin{align}
    I_{SO(\infty)}&=\text{PE}\left[\frac{2t^4(3+2t^2+t^4)}{(1-t^2)(1+t^2)^3}\right]=1+6t^4-8t^6 +39t^8-72t^{10}+ 230t^{12}+\cdots
\end{align}
The entropy $S(E)$ scales as $\sim E^{3/4}$ for large $E$. 
Interestingly, the indicial entropy of SO/Sp is greater than that of SU \eqref{SUinf}. This is because there is no cancellation between bosons and fermions in the index, as bosonic gravitons carry charge $j=4n$ and fermionic gravitons $j=4n+2$.

\subsection{SO(7) and Sp(3) BMN indices}\label{so7-so3}



One can  compute SO(7) and Sp(3) indices in closed form by summing over all residue contributions in \eqref{bmnsospind2}.
As in the previous section, we set equal fugacities \( a = b = c = t^2 \). 
The BMN indices are given as rational functions, expressed as the ratio of a numerator and a denominator.

The denominators of the BMN indices for SO(7) and Sp(3) share similarities but are different.
\begin{align}\label{difden}
    &\text{den}_{SO(7)} = \text{common factor} \times (1 - t^{20})^3 (1 - t^{22})^3, \nonumber\\
    &\text{den}_{Sp(3)} = \text{common factor} \times (1 - t^{12})^6.
\end{align}  
where the common factor is given by  
\begin{align}
    \left(1 - t^{36}\right)\left(1 - t^{32}\right)^3\left(1 - t^{28}\right)^6 \left(1 - t^{24}\right)^{10} \left(1 - t^{20}\right)^{12} \left(1 - t^{16}\right)^9  \left(1 - t^{12}\right)^7\left(1 - t^8\right)^6 \left(1 - t^4\right)^6.
\end{align}  

\begin{figure}
    \centering
    \includegraphics[width=1\linewidth]{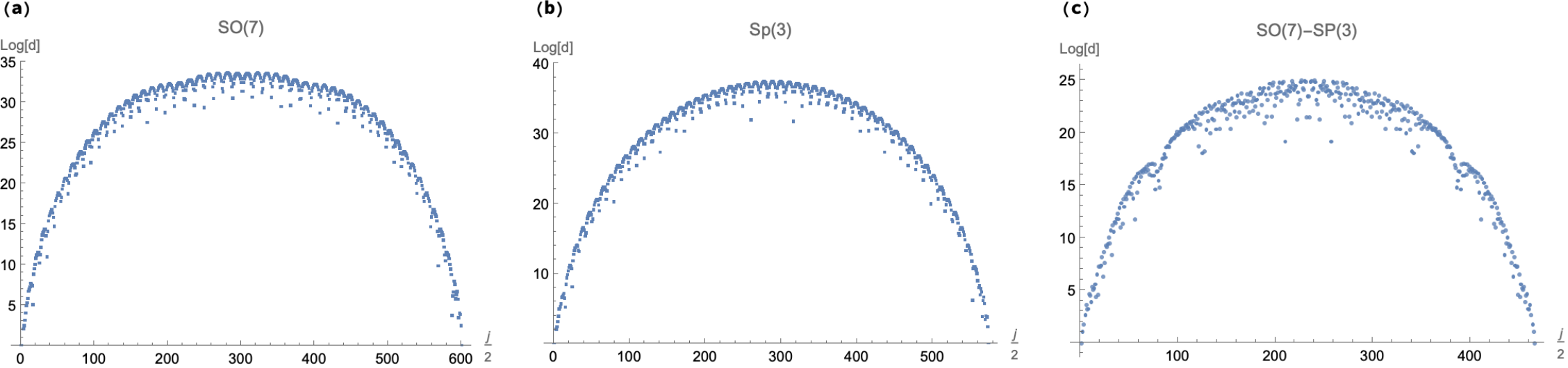}
    \caption{Logarithm of the absolute values of the coefficients in the numerators of the BMN indices for SO(7) and Sp(3).}
    \label{fig:sosp index}
\end{figure}

The numerators for the SO(7) and Sp(3) indices are given as follows. 
For SO(7), the numerator is  
\begin{align}  
    1 - 8t^6 + 12t^8 - 24t^{10} + 59t^{12} - 144t^{14} + 201t^{16} - 350t^{18} + \cdots + 55t^{1194} - 48t^{1196} + 12t^{1198} - t^{1200}.  
\end{align}  
For Sp(3), the numerator is  
\begin{align}  
    1 - 8t^6 + 12t^8 - 24t^{10} + 53t^{12} - 144t^{14} + 201t^{16} - 301t^{18} + \cdots - 48t^{1142} + 12t^{1144} - t^{1146}.  
\end{align}
The behavior of the coefficients in the numerators is illustrated in Figure \ref{fig:sosp index} (a) and (b), where we plot the logarithm of the absolute values of the coefficients, $\log |d(j)|$, against $j/2$. Recall,    $j = 2(Q_1 + Q_2 + Q_3) + 3(J_1 + J_2)$.

The difference between the  BMN indices is given by:  
\begin{align}
   I_{SO(7)} - I_{Sp(3)} &= \frac{-t^{18} (1 - t^2)^6}{(1 - t^8)^3 (1 - t^{16})^6 (1 - t^{12})^9(1 - 
   t^{20})^{15}(1 - t^{22})^3(1 - t^{24})^9(1 - t^{28})^6 } \nonumber\\
   &\quad \times (1+3t^2+6t^4-15t^8-39t^{10}-26t^{12}+\cdots+15t^{922}-6t^{926}-3t^{928}-t^{930}).
\end{align}  
In Figure \ref{fig:sosp index} (c), we plot $\log |d(j)|$ against $j/2$ for the coefficients of the polynomial  appearing in the numerator of this difference. It is interesting to note that these coefficients in the numerator exhibit an anti-palindromic symmetry: $d(j) = -d(930 - j)$. 

\begin{figure}
    \centering
    \includegraphics[width=0.5\linewidth]{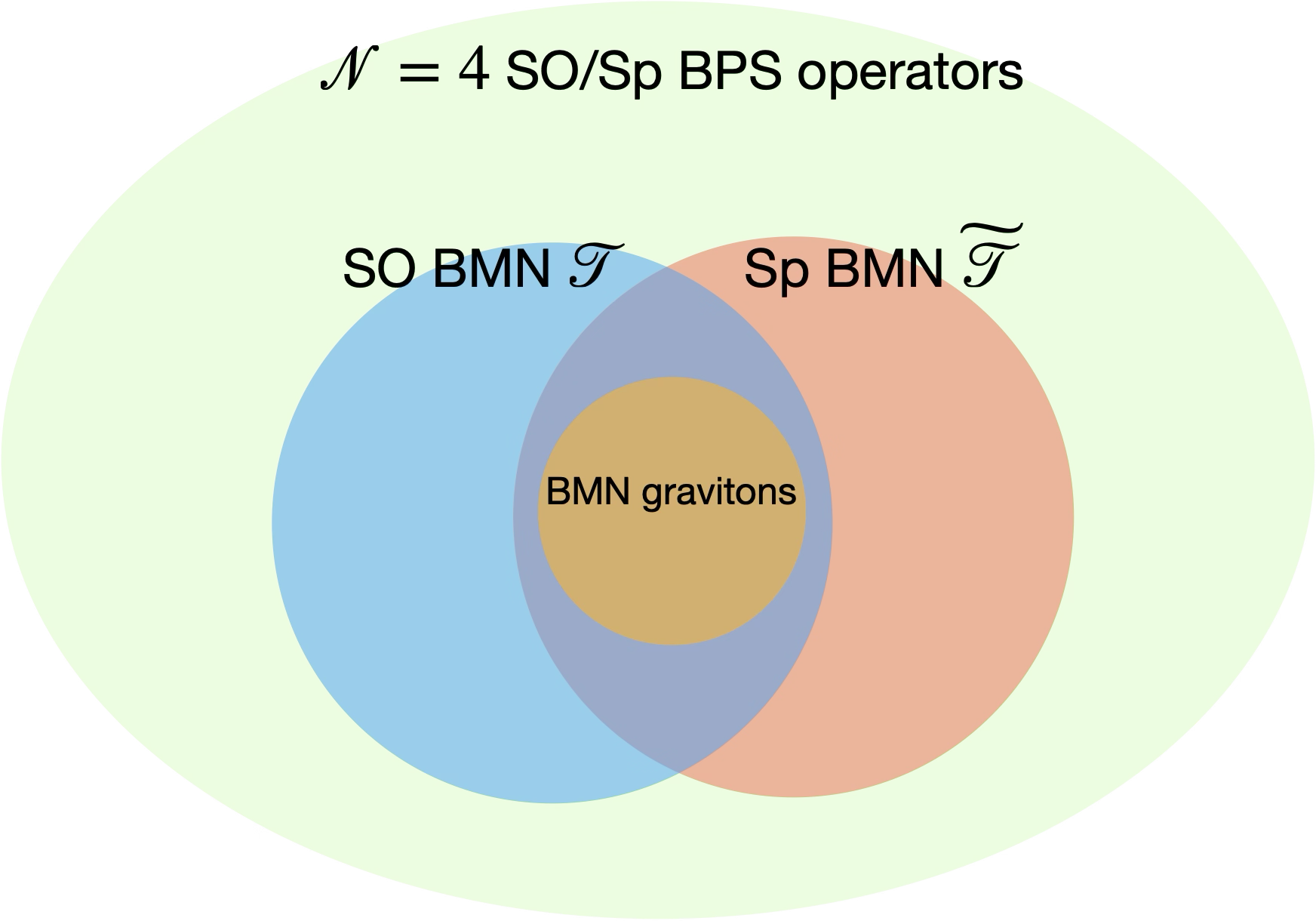}
    \caption{A schematic Venn diagram of the BMN sectors in SO($2N+1$) and Sp($N$) theories. The left circle represents SO BMN operators.  The right circle represents Sp BMN operators. Their overlap is non-empty, containing the graviton operators as a subset.}
    \label{fig:venn_BMN}
\end{figure}

Because the BMN graviton spectra for SO($2N+1$) and Sp($N$) coincide the difference between the two indices is attributed completely to non-graviton states.\footnote{As discussed in the beginning of this section, graviton cohomologies are described by products of \( S_n \) multiplets with \( n = \text{even} \), and the single gravitons in the BMN sector in both gauge theories are the same: \( |n\rangle \) and \( \bar{\mathcal{Q}}^m_{+} \bar{\mathcal{Q}}^n_{+}|n\rangle \).  
For \( SO(7) \) and \( Sp(3) \), the single graviton operators belong to the \( S_2 \), \( S_4 \), and \( S_6 \) multiplets.}
Figure \ref{fig:venn_BMN} schematically illustrates the Hilbert space of the SO/Sp BMN sectors.
We emphasize that this approach is useful since one can readily identify the subset of non-graviton operators by just computing the BMN indices of two dual theories, without directly computing the graviton index -- which is in general regarded as a difficult task for high-rank gauge groups.

\subsection{SO(7) non-graviton cohomology}
In this section we will track down the lightest non-graviton operator responsible for the mismatch between the two BMN indices. It appears at ${\cal O}(t^{18})$. Because $I_{SO(7)}-I_{Sp(3)}=-t^{18}+\ldots$, the operator could either be a fermionic operator in the SO(7) theory or a bosonic operator in the Sp(3) theory. This is settled by computing the graviton index. The graviton index of both the theories is the same and is given by  
\begin{align}
    I_{\text{grav}} = 1 + 6t^4 - 8t^6 + 39t^8 - 72t^{10} + 230t^{12} - 504t^{14} + 1203t^{16} - 2653t^{18} + \cdots.
\end{align}  
Comparing this result with the BMN indices given in the previous subsection, we find a discrepancy at order $t^{18}$ in the SO(7) theory. This confirms the existence of at least one fermionic non-graviton operator in SO(7).\footnote{It could be that SO(7) has two fermionic  and one bosonic non-graviton operators at this order and so on.}  Of course, by $S$-duality, there must also be a corresponding fermionic non-graviton operator in Sp(3) that lies outside the BMN sector.

Next we identify the charges of the non-graviton operators.    
Since the BPS letters have the following quantum numbers:  
$\bar{\phi}$ carries $j=2$, $\psi$ carries $j=4$, $f$ carries $j=6$,  
a general fermionic operator can be constructed from the following letter combinations:  
\begin{itemize}
    \item 8-letter operators: $(\psi, \bar{\phi}^7)$  
    \item 6-letter operators: $\left((\psi, \bar{\phi}^4, f), (\psi^3, \bar{\phi}^3) \right)$  
    \item 4-letter operators: $(\psi^3, f)$  
\end{itemize}  
If there is only one non-graviton operator then it is in a singlet representation of SU(3). Therefore, the non-graviton state has equal charges, i.e.,  $Q_1 = Q_2 = Q_3, ~ J_1 = J_2$. If this property were not true, we could have permuted these charges to obtain other non-graviton operators. 

The 1-loop Hamiltonian commutes with the bonus symmetry $U(1)_Y$ that counts the number of letters. In other words, the sectors of operators with different number of letters do not mix at 1-loop. Since the operator that we are looking for is a non-graviton operator. It must be in the $8$-letter sector because the trace relations appear only in the sector with more than $7$ letters. The operator constructed out of letters $(\psi, \bar \phi^7)$ has charges
\begin{align}
    (Q_1,Q_2, Q_3, J_1, J_2)=\Big(\frac52,\frac52,\frac52,\frac12,\frac12\Big)
\end{align}
which indeed obeys $Q_1 = Q_2 = Q_3, ~ J_1 = J_2$.

To identify the number of non-graviton operators in a given letter sector, we use the following decomposition:  
\begin{align}
    \text{Free BPS} &= \text{non-}\mathcal{Q}\text{-closed} \oplus \mathcal{Q}\text{-closed},\nonumber\\
    \mathcal{Q}\text{-closed} &= \mathcal{Q}\text{-cohomology} \oplus \mathcal{Q}\text{-exact},\\
    \mathcal{Q}\text{-cohomology} &= \text{graviton} \oplus \text{non-graviton}.\nonumber
\end{align}  
Rewriting this, we obtain:  
\begin{align}
    \text{Free BPS} = \text{non-}\mathcal{Q}\text{-closed} \oplus \mathcal{Q}\text{-exact} \oplus \text{graviton} \oplus \text{non-graviton}.
\end{align}  
Thus, to determine the number of non-graviton operators, we count the independent free BPS operators, the non-$\mathcal{Q}$-closed operators, the $\mathcal{\mathcal{Q}}$-exact operators, and the graviton operators.  
The total number of BPS operators in the free theory  in this $8$ letter sector with charges $(\frac52,\frac52,\frac52,\frac12,\frac12)$ is 903. 
By computing the $\mathcal{Q}$-action explicitly, we count:  
\begin{align}
    \text{Non-}\mathcal{Q}\text{-closed states: } & 220, \nonumber \\
    \mathcal{Q}\text{-exact states: } & 559, \nonumber \\
    \text{Graviton states: } & 123, \nonumber \\
    \text{Non-graviton states: } & 1.
\end{align}  
This confirms the existence of a single non-graviton cohomology at order $t^{18}$ in the SO(7) theory.
The explicit expression for the leading non-graviton $\mathcal{Q}$-cohomology is
\begin{align}
    &\mathcal{O}=\text{Tr}[Y^2] \text{Tr}[X\psi_1] \text{Tr}[XZ]^2-4 \text{Tr}[Y^2] \text{Tr}[XZ]
   \text{Tr}[ZXZ\psi_3]-\text{Tr}[XZ]^2
   \text{Tr}[ZY\psi_3Y]\nonumber\\
   &-4 \text{Tr}[XZ]^2
   \text{Tr}[ZY^2\psi_3]+8 \text{Tr}[XZ] \text{Tr}[ZXY^2Z\psi_3]+4
   \text{Tr}[XZ] \text{Tr}[ZXZY\psi_3Y]\nonumber\\
   &+16 \text{Tr}[XZ]
   \text{Tr}[ZXZY^2\psi_3]-4 \text{Tr}[Z\psi_3] \text{Tr}[ZX]
   \text{Tr}[ZXY^2]+8 \text{Tr}[ZXZ\psi_3] \text{Tr}[XYZY]\nonumber\\
   &-2 \text{Tr}[Y^2]
   \text{Tr}[Z\psi_3] \text{Tr}[ZXZX]+8 \text{Tr}[ZXZX]
   \text{Tr}[Y^2Z\psi_3]+2 \text{Tr}[ZXZX] \text{Tr}[YZY\psi_3]\nonumber\\
   &+16\text{Tr}[YZ\psi_3] \text{Tr}[ZXZXY]+8 \text{Tr}[Z\psi_3]
   \text{Tr}[ZXZXY^2]+8 \text{Tr}[Y^2] \text{Tr}[ZXZXZ\psi_3]\nonumber\\
   &+16\text{Tr}[ZXZXYZY\psi_3]-8 \text{Tr}[ZXZXZY\psi_3Y]-32
   \text{Tr}[ZXZXZY^2\psi_3]-16 \text{Tr}[ZXZYXYZ\psi_3]\nonumber\\
   &-16\text{Tr}[ZXZYXZY\psi_3]-16 \text{Tr}[ZXZY^2XZ\psi_3]-(grav)
\end{align}
Here, $-(grav)$ means that we project the state to be orthogonal to the graviton state vector space.
This can be done by projecting each term individually.
For example, after projection, the first term in the equation, $\text{Tr}[Y^2] \text{Tr}[X\psi_1] \text{Tr}[XZ]^2$, becomes $\text{Tr}[Y^2] (\frac{1}{3}\text{Tr}[X\psi_1+Y\psi_1+Z\psi_1]) \text{Tr}[XZ]^2$.
The expression is unique up to $\mathcal{Q}$-exact terms.
We confirmed that the operator is fortuitous -- it is not $\mathcal{Q}$-closed for SO($2N+1$) with $N > 3$.

\subsection{S-duality and testing the 1-loop nonrenormalization conjecture}
Does the existence of the 1-loop BPS operator  $ {\cal O}$ in $SO(7)$ theory with charges $(\frac52,\frac52,\frac52,\frac12,\frac12)$ signal the existence of 1-loop BPS operator of the same charge in $Sp(3)$ theory due to S-duality? This does not follow in the absense of conjecutre $1$. This gives us a way to test conjecture $1$. On the other hand, it is also true that the full superconformal index of $SO(7)$ and $Sp(3)$, calculated in the free theory and also at 1-loop match. How is this consistent with the non-existence of the 1-loop BPS operator with matching charges?

In computing the superconformal index, we only keep track of the charges that commute with the supercharge. In particular, operators with charges $(\frac52+n,\frac52+n,\frac52+n,\frac12-3 n,\frac12-3 n)$ contributes the same to the superconformal index for any value of $n$ -- in particular same as operator ${\cal O}$ ($n=0$). Hence S-duality only  implies the existence of an operator $\tilde {\cal O}$ at 1-loop  in the $Sp(3)$ theory with the above charges for some value of $n$.
In fact because the superconformal index counts with $(-1)^F$, it predicts that there must be one extra fermionic BPS operator as compared to bosonic with charges of the type $(\frac52+n,\frac52+n,\frac52+n,\frac12- n,\frac12- n)$. 

If we do not find precisely one fermionic BPS operator in the Sp(3) theory with charges $(\frac52,\frac52,\frac52,\frac12,\frac12)$ then conjecture $1$ must be false. The operator with those charges that is not in the BMN sector has two possibilities for its letter content: $(\bar{\lambda}^2,\psi, \phi^4)$ and $(D, \bar{\lambda}, \phi^6)$. We leave the investigation of whether the operator exists to future work. 
Of course we can also analyze the 1-loop BPS spectrum at higher charges and check whether they match as well. As shown in Eq.~\eqref{difden}, the tower structures of the SO(7) and Sp(3) BMN indices differ. For instance, the SO(7) BMN index contains \(t^{20}\) towers refined by fugacities \(a^4b^4c^2\) and its permutations. A natural bosonic contribution with this structure is operators like \(f^2\phi_1^2\phi_2^2\) (and its permutations). By \(S\)‑duality, there must be  towers of operators with matching charges in the Sp(3) theory outside the BMN sector.
It would be interesting to know the detailed tower structures of SO and Sp.

As remarked earlier, conjecture $2$ is strictly stronger than conjecture $1$. So even if conjecture $1$ holds, we can probe the validity of conjecture $2$. Conjecture $2$ implies the existence of a  sub-ring $\tilde {\cal T}$ for $Sp(3)$ that is dual to  the sub-ring $\cal T$ for $SO(7)$. For the BMN truncation, if the operator ${\tilde {\cal O}}$ matching with operator ${\cal O}$ does indeed exist in the $Sp(3)$ theory then the conjecture $2$ predicts that multiplying it by any of the low-lying BMN cohomology element must result in a cohomology element that is dual to the cohomology obtained by taking the product of ${\cal O}$ and the corresponding low-lying BMN cohomology in $SO(7)$ theory. If this does not turn out to be the case then conjecture $2$ is false. It is natural to explore this possibility if conjecture $1$ turns out to be true.

\subsection{Other letter-based truncations}\label{1ltrun}

As we have seen in the BMN 1-loop truncation, truncating the full 1/16-BPS sector is a useful strategy for identifying non-graviton 1-loop cohomologies, as it reduces computational cost and facilitates analytic analysis of the index structure.  
In particular, since the 1-loop truncated indices of  SO  and  Sp  theories do not necessarily match, their difference -- when considered alongside the \( S \)-duality of the full theories -- provides direct evidence for the existence of non-graviton cohomologies.
In this subsection, we apply the same strategy that we used for the BMN truncation to  another non-trivial 1-loop truncation. We observe the mismatch between SO and Sp indices for  this sector, and identify the  charges of the non-graviton operator at $j=22$.

We consider the sector consisting  of the following letters:  
\begin{align}
    X, \bar{\lambda}_{\dot{\alpha}}, D_{+\dot{\alpha}}.
\end{align}  
Its single-letter index is given by  
\begin{align}
    f = 1 - \frac{(1-a)}{(1-p)(1-q)}.
\end{align}  
The $\mathcal{Q}$-action on these letters is given by  
\begin{align}
  &&[\mathcal{Q},X] = 0, \quad \{\mathcal{Q},\bar\lambda_{\dot\alpha}\} = 0 ,\quad 
  [\mathcal{Q},D_{+\dot\alpha}](\cdots) = -i g_{\rm YM} [\bar\lambda_{\dot{\alpha}},(\cdots)\}.
\end{align}  
The algebra closes within this subsector as expected.  
To investigate whether this subsector contains non-graviton cohomologies, we compute the indices for SO($2N+1$) and Sp($N$).  
The SO($2N+1$) and Sp($N$) index in this sector is given by the following integrals respectively.
\begin{align}
I_{SO(2N+1)}=&\frac{(p;p)_{\infty}^N(q;q)_{\infty}^N}{2^N N!}\left(\text{PE}\left[\frac{a-pq}{(1-p)(1-q)}\right]\right)^N \nonumber\\
&\oint \prod_{i=1}^{N} \frac{dz_i}{2\pi i z_i} \text{PE}\left[\frac{a-1}{(1-p)(1-q)}\Big(\sum_{i<j}\Big((\frac{z_i}{z_j})^\pm + (z_i z_j)^{\pm}\Big)+\sum_{i} z_i^{\pm}\Big)\right]\\
I_{Sp(N))}=&\frac{(p;p)_{\infty}^N(q;q)_{\infty}^N}{2^N N!}\left(\text{PE}\left[\frac{a-pq}{(1-p)(1-q)}\right]\right)^N \nonumber\\
&\oint \prod_{i=1}^{N} \frac{dz_i}{2\pi i z_i} \text{PE}\left[\frac{a-1}{(1-p)(1-q)}\Big(\sum_{i<j}\Big((\frac{z_i}{z_j})^\pm + (z_i z_j)^{\pm}\Big)+\sum_{i} z_i^{\pm2}\Big)\right]
\end{align}

Taking $a=t^2$, $p=q=t^3$ and expanding both the integrals for $N=3$ we get,
\begin{align}
    I_{SO(7)}=1 &+ t^4 - 2 t^5 + t^6 + 2 t^7 - 2 t^8 - 2 t^9 + 6 t^{10} - 4 t^{11} - 
 5 t^{12} + 10 t^{13} + 2 t^{14} - 18 t^{15} \nonumber\\
 &+ 15 t^{16}+ 10 t^{17} - 33 t^{18} + 
 14 t^{19} + 40 t^{20} - 56 t^{21} - 4 t^{22}+84 t^{23}\cdots\\
 I_{Sp(3)}=1 &+ t^4 - 2 t^5 + t^6 + 2 t^7 - 2 t^8 - 2 t^9 + 6 t^{10} - 4 t^{11} - 
 5 t^{12} + 10 t^{13} + 2 t^{14} - 18 t^{15} \nonumber\\
 &+ 15 t^{16}+ 10 t^{17} - 33 t^{18} + 
 14 t^{19} + 40 t^{20} - 56 t^{21} - 5 t^{22}+86t^{23}+\cdots
\end{align}
The SO/Sp  indices match up to \( t^{21} \), but a mismatch appears at \( t^{22} \). This mismatch indicates the presence of fortuitous states with refined fugacities \( x^2 p^3 q^3 \), corresponding to either a bosonic state in  SO(7)  or a fermionic state in  Sp(3) .  

One can estimate the entropy growth of the non-graviton states in the truncated sector for large $N$, using a saddle point analysis in the Cardy-like limit, where \( p, q \to 1 \). This is equivalent to taking the limits \( \omega_{1,2} \to 0 \) with \( \text{Re} \, \omega_{1,2} > 0 \), where \( p = e^{-\omega_1} \) and \( q = e^{-\omega_2} \). The eigenvalues condense into a delta-function distribution:  
\[
\rho(\alpha) = \delta(\alpha).
\]  
Since $1-p\approx \omega_1,~1-q\approx \omega_2$, the free energy is given by  
\begin{align}
\log I\approx N^2 \sum_{n=1}^{\infty} \frac{e^{-n\Delta} - 1}{n^3 \omega_1 \omega_2} = N^2 \frac{\text{Li}_3(e^{-\Delta}) - \zeta(3)}{\omega_1 \omega_2}.
\end{align}  
Note that we computed the free energy in the large‑$N$ limit for the SU($N$) theory. For SO($2N+1$) and Sp($N$), the free energy is doubled, and the corresponding entropy is halved compared to that of SU($N$).  

The entropy is obtained via the inverse Laplace transform:  
\begin{align}
S(J_1, J_2, Q) = \text{Ext}_{\omega_1,\omega_2,\Delta} \left[ N^2 \frac{\text{Li}_3(e^{-\Delta}) - \zeta(3)}{\omega_1\omega_2} + \omega_1 (J_1 + Q) + \omega_2 (J_2 + Q) + \Delta Q_1 \right].
\end{align}
where $Q=\frac{Q_1+Q_2+Q_3}{3}$
\begin{align}
J_1 + Q = N^2 \frac{\text{Li}_3(e^{-\Delta}) - \zeta(3)}{\omega_1^2 \omega_2}, \quad J_2 + Q = N^2 \frac{\text{Li}_3(e^{-\Delta}) - \zeta(3)}{\omega_1 \omega_2^2}, \quad Q_1 = N^2 \frac{\text{Li}_2(e^{-\Delta})}{\omega_1\omega_2}.
\end{align} 
Requiring the charges to be real and positive leads to the chemical potentials being complex.  

Comparison to the full entropy: 
BPS black hole entropy in the full sector is given as
\begin{align}
    S_{BH}=2\pi N^2\sqrt{q_1q_2+q_2q_3+q_3q_1-(j_1+j_2)/2}
\end{align}
where $q_i=\frac{Q_i}{N^2}$, $j_a=\frac{J_a}{N^2}$.
BPS black holes live on a codimension-1 surface in the five-dimensional charge space, satisfying the following non-linear charge relation:
\begin{align}
    q_1q_2+q_2q_3+q_3q_1-(j_1+j_2)/2=\frac{q_1q_2q_3+j_1j_2/2}{q_1+q_2+q_3+1/2}
\end{align}
For simplicity let us assume $\omega_1=\omega_2$ and $\Delta_2=\Delta_3$ so that $J_1=J_2=J$ and $Q_2=Q_3$. In the Cardy limit, 
\begin{align}
    (2q_1q_2+q_1^2)(q_1+2q_2)-q_1q_2^2\approx j^2/2,
\end{align}
where $j$ is of order $O(q^{3/2})$.

Although we do not have an analytic proof, it seems that the entropy of the truncated sector is smaller for arbitrary charges. For example, for $j\approx 3q_1^{3/2}$,
(ex: $j\sim 10^9$, $q_1\sim \left(\frac{1}{3}\right)^{2/3}10^6$, $q_2=2.897 \times 10^5$)
 $S\sim4.48\times 10^6$.
For the truncated sector with the same charges, $\text{Re}[S]\sim 2.32\times 10^6$.

\section{Discussion}\label{discussion}

In this paper, we analyze spectral structures of non-graviton cohomologies.
We have mostly restricted our attention to 1-loop $\mathcal{Q}$-cohomologies in the BMN sector, since it reduces computational complexity yet contains order-$N^2$ non-graviton states.

In Section \ref{bmntr}, we suggest an algorithm to obtain BMN indices for SU($N$) $\mathcal{N}=4$ SYM.
We showed that the BMN index is given by a rational function, whose denominator structure provides direct evidence for the existence of non-graviton states.
Furthermore, we find that there are tower structures for a certain class of non-graviton operators. It would be interesting to understand the gravity dual interpretation of this tower.

In Section \ref{sospbmn}, we discuss the S-duality between SO($2N+1$) and Sp($N$) $\mathcal{N}=4$ SYM.
Although the theories are S-dual to each other, their BMN indices are not manifestly S-dual.
The fact that various 1-loop truncated indices are not S-dual allows us to identify certain non-graviton states even without computing graviton indices. 
Using the strategy, we have identified one of SO(7) non-graviton cohomologies. We have also identified research directions that could potentially falsify the conjecture.  We would like to purse these directions in the future. In case either of the conjecture is false, it would be interesting to find out if it is the higher order perturbative effects or the non-perturbative effects that  play the spoil-sport. We would also like to have a systematic and effecient way to compute graviton BMN index for higher rank gauge groups so that our analysis can be pushed to higher rank.

Interestingly, the emergence of a new symmetry at the 1-loop level plays a crucial role in our analysis, as it enables a consistent truncation to the BMN sector.
However, its interpretation from the gravity side remains unclear. Since this is a symmetry of the 1-loop theory, and as we have seen in Section \ref{sospbmn}, it appears to be broken at strong coupling, suggesting that it may not have a meaningful counterpart in the full non-perturbative regime.
Nevertheless, it is intriguing to speculate whether this symmetry is related in some way to the U(1) bonus symmetry observed in supergravity, as discussed by Intriligator in \cite{Intriligator:1998ig}.

Using the idea of \cite{Chen:2022hbi}, the authors of \cite{Choi:2022asl} claimed that it is possible to compute the spectral form factor using the supersymmetric index, capturing characteristics of black hole spectra. 
It would be interesting to see whether the exact expression of the BMN index can be used to compute the spectral form factor, and to explore diagnostics of chaotic spectra, as discussed in \cite{Chen:2024oqv}.
There appears to be a universal shape in the distribution of coefficients in the numerator of the BMN index (see Figures \ref{fig:bmn index coeff} and \ref{fig:sosp index}).
It would be desirable to find a quantitative characterization of this shape.
We also observed structural patterns in the denominator but did not study them in full generality.
If the general form of both the numerator and denominator is determined, one may be able to extract the universal entropy growth of the BMN sector.

Finally, it would be interesting to explore 1-loop nonrenormalization conjecture in theories with ${\cal N}=2$ that exhibit S-duality such as SU(2) gauge theory with four flavors. It would also be interesting to explore the non-graviton spectrum in  theories with lower supersymmetry.

\section*{Acknowledgement}
We would like to thank Chi-Ming Chang, Jaehyeok Choi, Seok Kim, Shiraz Minwalla, Chintan Patel for very useful comments and discussions. 
The work of E.L. was supported by Basic Science Research Program through the National Research Foundation of Korea (NRF) funded by the Ministry of Education RS-2024-00405516.
We would like to acknowledge the support of the Department of Atomic Energy, Government of India, under Project Identification No. RTI 4002. This work is partially supported by the Infosys Endowment for the study of the
Quantum Structure of Spacetime. The authors would also like to acknowledge their debt to the people of India for their steady support to the study of the basic sciences.

\appendix

\bibliography{biblio.bib}

\end{document}